\documentclass[12pt,epsfig]{article}

\usepackage{multirow}
\usepackage[dvips]{graphics}
\usepackage{rotating}
\usepackage{epsfig}
\def\lesssim{\mathrel{\hbox{\rlap{\hbox{\lower5pt\hbox{$\sim$}}}\hbox{$<$}}}}
\def\gtrsim{\mathrel{\hbox{\rlap{\hbox{\lower5pt\hbox{$\sim$}}}\hbox{$>$}}}}

\newcommand{\ntrl}[1]{\chi^0_#1}
\newcommand{\chpm}[1]{\chi^\pm_#1}

\def\squark{\tilde{q}}

\newcommand{\sbot}[1]{\tilde{b}_#1}
\newcommand{\sstop}[1]{\tilde{t}_#1}
\def\squarkc{\tilde{q}^*}
\def\gluino{\tilde{g}}

\def\hpm{H^\pm}

\def\mone{M_1}
\def\mtwo{M_2}

\newcommand{\mntrl}[1]{m_{\chi^0_#1}}      %
\newcommand{\mchpm}[1]{m_{\chi^\pm_#1}}

\def\mslep{m_{\tilde{\ell}}}

\def\msquark{m_{\tilde{q}}}

\def\mgluino{m_{\tilde{g}}}

\def\mhpm{m_{H^\pm}}

\def\tanbeta{\tan\beta}

\def\beq{\begin{equation}}   %
\def\eeq{\end{equation}}   %

\begin{document}

\begin{flushright}
   {\bf HRI-P-08-06-001 \\
 RECAPP-HRI-2008-007}
\end{flushright}

\vskip 30pt

\begin{center}
{\large \bf Signatures of gaugino mass non-universality in cascade
Higgs production at the LHC}\\
\vskip 20pt
{Priyotosh Bandyopadhyay\footnote{priyotosh@mri.ernet.in},
AseshKrishna Datta\footnote{asesh@mri.ernet.in} \\ and 
Biswarup Mukhopadhyaya\footnote{biswarup@mri.ernet.in}}  \\
\vskip 20pt
{ \emph{Regional Centre for Accelerator-based Particle Physics  \\
Harish-Chandra Research Institute  \\
Chhatnag Road, Jhunsi, Allahabad, India 211019}}\\

\end{center}

\vskip 65pt

\abstract{}
Supersymmetric cascades, involving charginos and neutralinos at
various stages, contribute in a significant way
to Higgs production at the LHC. We explore the nature of such
cascades, completely relaxing the universality of the gaugino masses.
 It is found that the deviation from the scenario with universal
gaugino  masses would be reflected in the relative production rates for the
lightest Higgs and the charged Higgses, two characteristic
 particles of an extended Higgs sector.

\newpage 

\section{Introduction}
Supersymmetry (SUSY) stands out as one of the most interesting
alternatives beyond the Standard model (SM) of elementary particles. The Minimal Supersymmetric
extension of the Standard Model (MSSM) 
necessarily contains two Higgs doublets, which, on electroweak symmetry
breaking (EWSB), leads to 5 physical Higgs bosons, namely, two CP-even neutral 
scalars ($h, H$), one CP-odd Higgs ($A$) and two mutually conjugates
charged scalars ($H^\pm$). In such a framework, the Higgs phenomenology is
obviously richer than in the SM \cite{Djouadi:2005gi,AguilarSaavedra:2005pw}. With the Large Hadron Collider (LHC)
all set to take off, the hunt for this yet undiscovered scalar sector 
has assumed a spacial significance, concurrently with the search for SUSY.

The phenomenology of the Higgs sector in the MSSM is considerably
enriched by the interaction of the various scalar (Higgs) states 
with SUSY particles \cite{Haber:1984rc,HHG,Gunion:1984yn}. Notwithstanding the prevailing emphasis on
Higgs production in processes driven by SM interactions, the prospect
of extracting information from SUSY channels and cascades should
therefore be always kept within sight. As for example, viable Higgs
signals in associated production of  superparticle have been suggested
in recent studies, for cases where SM channels fail due to the
effects of CP-violating phases \cite{Bandyopadhyay:2007cp}. In this paper, we suggest the
utilisation of Higgs production in SUSY cascades in probing the
chargino-neutralino sector of the MSSM. In particular, we show that
the relative rates of $h$-and $H^{\pm}$-production in cascades can
provide insight on whether $M_1$ and $M_2$, the $U(1)$ and $SU(2)$ gaugino
masses, respectively,  are related by a high-scale universality condition.

The significance of SUSY cascades as the source of the MSSM 
Higgs bosons has  been discussed in detail in the recent 
past \cite{Datta:2003iz} within an MSSM
framework, but keeping $M_1$ and $M_2$ constrained by universality. 
The central idea in such a study has been to exploit the huge 
production cross sections for the strongly interacting SUSY 
particles the squarks and the gluino  
at a hadron collider like the LHC.
These sparticles, once produced, may undergo long cascade-decays that
ultimately lead to stable SM particles (like leptons and quarks (jets))
along with the LSP's which escape detection if R-parity (defined
as $R~=~(-)^{3B+L+2J}$) is conserved.  It was pointed out in 
\cite{Datta:2003iz,Baer:1992ef}that one or 
more MSSM Higgs bosons can be produced at different stages of these 
cascades. It was also shown \cite{Datta:2001qs} that the suppressions 
resulting from different decay
branching fractions under SUSY cascades are more than compensated for by
the huge production cross-section of the strongly interacting particles.
It is to be noted that such cascades, in order to be instrumental in
Higgs production, necessarily require SUSY particles widely separated
in mass. A high energy machine like LHC is an
ideal hunting ground for MSSM Higgs bosons under such cascades.

As has been already pointed out in the earlier works 
\cite{Datta:2001qs}, cascades can be very efficient sources of MSSM 
Higgs bosons in certain regions of MSSM parameter space (viz., with 
intermediate $\tanbeta$ values) where usual modes cease to deliver.
Here we make use of these sources, with the $M_1 - M_2$ universality
conditions relaxed, and discover some rather spectacular consequences
on the overall rates.

In canonical SUSY scenarios, the neutralinos and the charginos (which are 
the mass eigenstates and mixtures of electroweak gauginos and the higgsinos)
can be much lighter compared to the strongly interacting sparticles like 
squarks and gluinos. Thus, charginos and neutralinos may  take control of the 
proceedings at an early stage of the cascade. Further, their
compositions (in terms of the gaugino and Higgsino contents), which play a
crucial role. The compositions in turn are determined by the soft
masses of the electroweak 
gauginos, namely, $\mone$ and $\mtwo$, $\mu$,
the so-called higgsino mass parameter, and 
$\tanbeta$, the ratio of the vacuum expectation values of the two Higgs
doublets. In particular, the relative magnitudes of $\mone$, 
$\mtwo$ and $\mu$ largely
determine their physical states.

There is no {\it a priori} justification as to why gaugino
universality, albeit highly predictive and hence, popular, 
would necessarily hold at a high scale. In fact, it has its root 
in the trivial nature of the so-called gauge kinetic function from 
which the common gaugino  mass arises at a high scale as SUSY breaks 
in the hidden sector. More specifically, such a universality 
arises when the gauge kinetic function involves a 
combination of the hidden sector fields which is singlet 
under the underlying gauge group of SUSY Grand Unified Theory(GUT).  It has been 
shown [\cite{Martin}-\cite{Cremmer}] that involving higher GUT representations for the purpose 
would in general trigger nonuniversality among the gaugino soft masses at a 
high scale itself. Such a nonuniversality inevitably distorts the 
weak-scale gaugino spectrum thus modifying the compositions of the 
charginos and the neutralinos and their masses vis-a-vis the
gluino mass. From a purely phenomenological point of view one can thus
think of a completely uncorrelated gaugino sector at the weak scale. 
This can have profound implications
in collider data [\cite{Bt1}-\cite{Huitu:2007vw}]. It is 
interesting to note that imprints of such nonuniversality can be recognised 
even in SUSY-Higgs searches at the LHC. Also, unlike in earlier works
\cite{Datta:2003iz,Datta:2001qs} we keep sleptons light enough so that
they have a nontrivial role to play.
 
The main consequence of relaxing the universality condition
on the $SU(3)$ gaugino mass ($M_3$) \cite{Datta:2003iz,Datta:2001qs} is that it gives a free hold to the 
gluino mass (and, in schemes of scalar mass evolution, the squark
masses). This affects the rates of cascades through gluino and
squark decay branching ratios only. By relaxing $M_1 - M_2$
universality, on the other hand, one opens up additional
possibilities, as far as the cascade branching ratios of the
charginos and neutralinos themselves are concerned. In addition, 
the lack of correlation between $M_1$ and $M_2$ affects the coupling
strengths of a charged or neutral Higgs to a chargino-neutralino
pair. Since such effects have not been studied systematically so far, 
we present an analysis here, in the context of the LHC.

In section 2 we outline the Higgs production process in cascades and
the factors that control them. In section 3 we the production rates of the 
charged $H^\pm$ and the lightest neutral Higgs 
($h$) bosons and contrast them systematically. We demonstrate how
such a knowledge could reflect on the nonuniversality of gaugino-masses 
We conlclude in section 4.

\section{Higgs production in SUSY cascades}
The squarks and the gluinos, once produced at LHC, would first undergo
strong two-body decays like $\squark \to  q \gluino$ (for $\msquark > \mgluino$) 
or $\gluino \to q \squark$ (for $\mgluino > \msquark$). Beyond this point, the  
cascade decays are electroweak in nature where Higgs bosons could appear 
\footnote{A possible exception could be when all squarks except the 
ones from the third
generation ($\sstop1$ or $\sbot1$) are heavier than the gluino. In such scenarios,
a cascade of strong decays  of squarks and gluinos might end up with $\sstop1$ 
or $\sbot1$ whose electroweak decays would lead to the Higgs bosons.}. 

Higgs production under such cascades mainly involves the 
charginos and neutralinos in the intermediate stages. With gluinos
initiating a cascade, this is inevitable, since gluinos do not couple
 directly to the Higgs bosons at the tree level.
For squarks, couplings to Higgs bosons are proportional to the corresponding quark 
masses, and are thus significant only for the squarks of the third family.
Since the generic yield of such squarks is smaller in comparison to those
of the first 
two families, most Higgs production processes in cascades involve
the charginos and neutralinos in the intermediate stages. 

Schematically, the chains 
of cascades leading to the Higgs bosons
are as follows:
\begin{equation}
pp \to \gluino \gluino,\; \squark \squark , \; \squark \squarkc, \; \squark \gluino
\longrightarrow \; \chpm2, \; \ntrl3, \; \ntrl4 \, + \, X \longrightarrow \; 
\chpm1, \; \ntrl2, \; \ntrl1 \, + \, \hpm, h, H, A \, + \, X  
\end{equation}
\vskip -15pt
\begin{equation}
pp \to \gluino \gluino,\; \squark \squark, \; \squark \squarkc, \; \squark \gluino
\longrightarrow \; \chpm1, \; \ntrl2 \, + \, X \longrightarrow \; 
\ntrl1,  \, + \, \hpm, \, h, H, A,  \hpm \, + \, X
\end{equation}
The first decay chain above is a longer one as it involves direct
decays of squark and gluinos to the heavier chargino/neuralinos followed by 
subsequent decays of the latter ones to lighter gauginos
and the Higgs bosons. 
On the other hand,  the second chain is shorter since it exploits
direct decays of squarks and gluinos to the lighter
chargino/neutralinos which then decay to Higgs bosons
and the LSP. In the literature \cite{Datta:2001qs} the first scheme was
called the `big cascade' while the latter one was dubbed as the 
`little cascade'. For convenience, we 
adopt the same terminology in this work.

It is thus expected that the final yield of Higgs bosons under such SUSY-cascades crucially
depends upon the branching fractions of the relevant decay
processes. Thus on  a complicated, though comprehensible, interplay of 
different SUSY parameters in the form of various couplings and masses. Out of these,
the couplings of the Higgs bosons with the charginos/neutralinos play
the most important role. It is well known \cite{Gunion:1984yn,Datta:2001qs}
that the Higgs bosons couple
favourably to charginos and neutralinos when the latter are mixtures of gauginos and
Higgsinos while for gauge bosons the couplings are maximal when the charginos and the
neutralinos are Higgsino-dominated. 

Naturally, then, the compositions of the charginos and neutralinos would play a crucial role
in our study \footnote{In scenarios with a universal gaugino mass  at a high scale 
(like the GUT scale), $\mone$ and $\mtwo$ gets related at the weak scale by the simple 
relation 
$\mtwo\simeq 2 \mone$. Thus, in that case, one only talks about 3 input parameters that govern
the chargino-neutralino sector. In contrast, the present work addresses the issue of 
nonuniversality of gaugino masses in a particular context. Hence, $\mone$ and $\mtwo$
are taken to be two free parameters.}. Out of the determining parameters, the values
of $\mone$, $\mtwo$ and $\mu$  have the most crucial bearings on the masses and the
contents of the charginos and the neutralinos. For $\mu >> \mone, \mtwo$ , one is in the
so-called `gaugino region' where the lighter neutralinos and chargino 
($\ntrl1,\ntrl2,\chpm1$) are 
gaugino-dominated with $\mntrl1 \simeq min(\mone, \mtwo)$ and 
$\mntrl2, \mchpm1 \simeq max( \mone,\mtwo)$
while the heavier ones ($\ntrl3, \ntrl4, \chpm2$) are mostly Higgsinos with
$\mntrl3, \mntrl4, \mchpm2 \simeq \mu$.  On the other hand,  
for $\mu << \mone, \mtwo$, we are in the `Higgsino region' for which the lighter neutralinos
and the chargino are predominantly Higgsinos with 
$\mntrl1, \mntrl2, \mchpm1 \simeq \mu$  while the heavier ones are dominated by
gauginos with 
$\ntrl3 \simeq min(\mone, \mtwo) $ and $\ntrl4, \mchpm2 \simeq max(\mone, \mtwo)$.
As expected, for different charginos and neutralinos, 
the masses and the contents have one-to-one correspondences in such `pure' regions of the
SUSY parameter space. For $\mone,\mtwo \simeq \mu$, the charginos and the neutralinos become maximally
mixed in gauginos and Higgsinos with their masses showing no particular pattern, albeit
restricted within a range determined by the values of $\mone, \mtwo$ and $\mu$.

In a nutshell, the `big cascades' are favoured in regions where, between 
($\chi_2^{\pm}, \chi_3^0, \chi_4^0$) and ($\chi_1^{\pm}, \chi_1^0, \chi_2^0$),
one set is gaugino-dominated, and the other, Higgsino-dominated. In
situations where they are kinematically allowed, little cascades are
on the hand possible when the members of the second set above
comparable gaugino and Higgsino components.

We investigate charged as well as neutral Higgs production rates in cascades.
For a ready comparison, we closely follow the earlier analyses 
\cite{Datta:2003iz,Datta:2001qs}. Explicit
expressions for most cross-sections and  decay widths of
relevance are found in the above references.

\section{The results of nonuniversality: numerical results}

We are looking at `effective cross-sections' of Higgs production of
various kinds, which essentially means 
$$
 pp \longrightarrow Higgs + X
$$
\noindent
where the cross-sections of all possible (from $2 \to 2$ strong
productions) cascades are added up, so
long as there is at least one Higgs of any kind in the final state.
 As cross-checks of the calculation, we have  reproduced the results 
in \cite{Datta:2003iz,Datta:2001qs} in the appropriate limits. We have used PYTHIA
\cite{Sjostrand:2006za} for our analysis and CTEQ3L \cite{CTEQ}
as parton distributions interfaced via LHAPDF \cite{CTEQ1}. The factorization/renormalization scale
set at the average of the masses of the particles (squarks and/or
gluinos) produced in the hard scattering. the analysis is based on
leading order production only. Also, following
\cite{Datta:2001qs}, the set of relevant  SM and SUSY inputs (at the weak scale) chosen 
for the analysis (unless otherwise specified) is:
\[  m_{top}=175 \ {\mathrm {GeV}} \quad   \tanbeta=10 \quad
m_A=162.1 (237.5) \ \mathrm{GeV} \] 
\[ \mgluino=900 \ \mathrm{GeV} \quad \msquark=800 \ 
\mathrm{GeV} \quad A_f=0    \]
This resulted in the following masses for the different Higgs bosons: 
\[ m_h=109 (110) \ \mathrm{GeV} \quad m_H=164 (238) \ \mathrm{GeV}
\quad m_{\hpm}=180 (250) \
\mathrm{GeV}                \]
which are in close agreement with those used earlier in the literature
\cite{Datta:2003iz,Datta:2001qs}. As for the sleptons, the only way they may signficantly contribute is by affecting the decay modes of 
the gauginos. As indicated in section 1, we demonstrated the role of
light sleptons, taking them to be degenerate at 400 GeV \footnote{We have checked that the use of more recent parton distribution 
functions, the variations in rates with renormalization/factorization 
scale and the use of an updated top-quark mass do not alter the
basic findings of the present work.}.

The parameter-dependence of Higgs production rates under SUSY cascades is investigated in two ways: (i) variation with
$\mtwo$ and (ii) variation with $\mu$.

\subsection{Variation with $\mtwo$}
\begin{figure}[hbt]
\begin{center}
{\epsfig{file=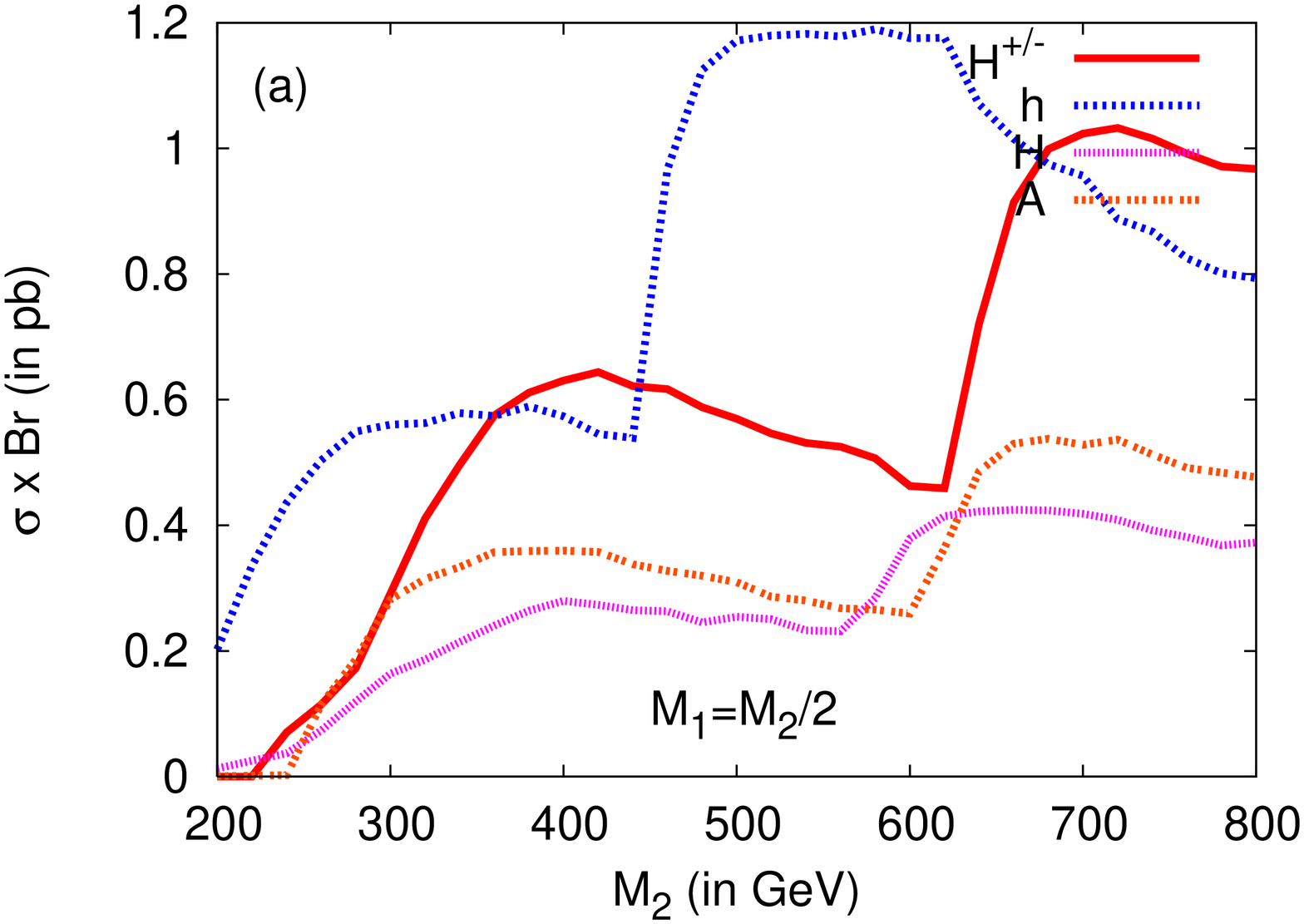,width=4.7 cm,height=5.2cm,angle=-90.0}}
{\epsfig{file=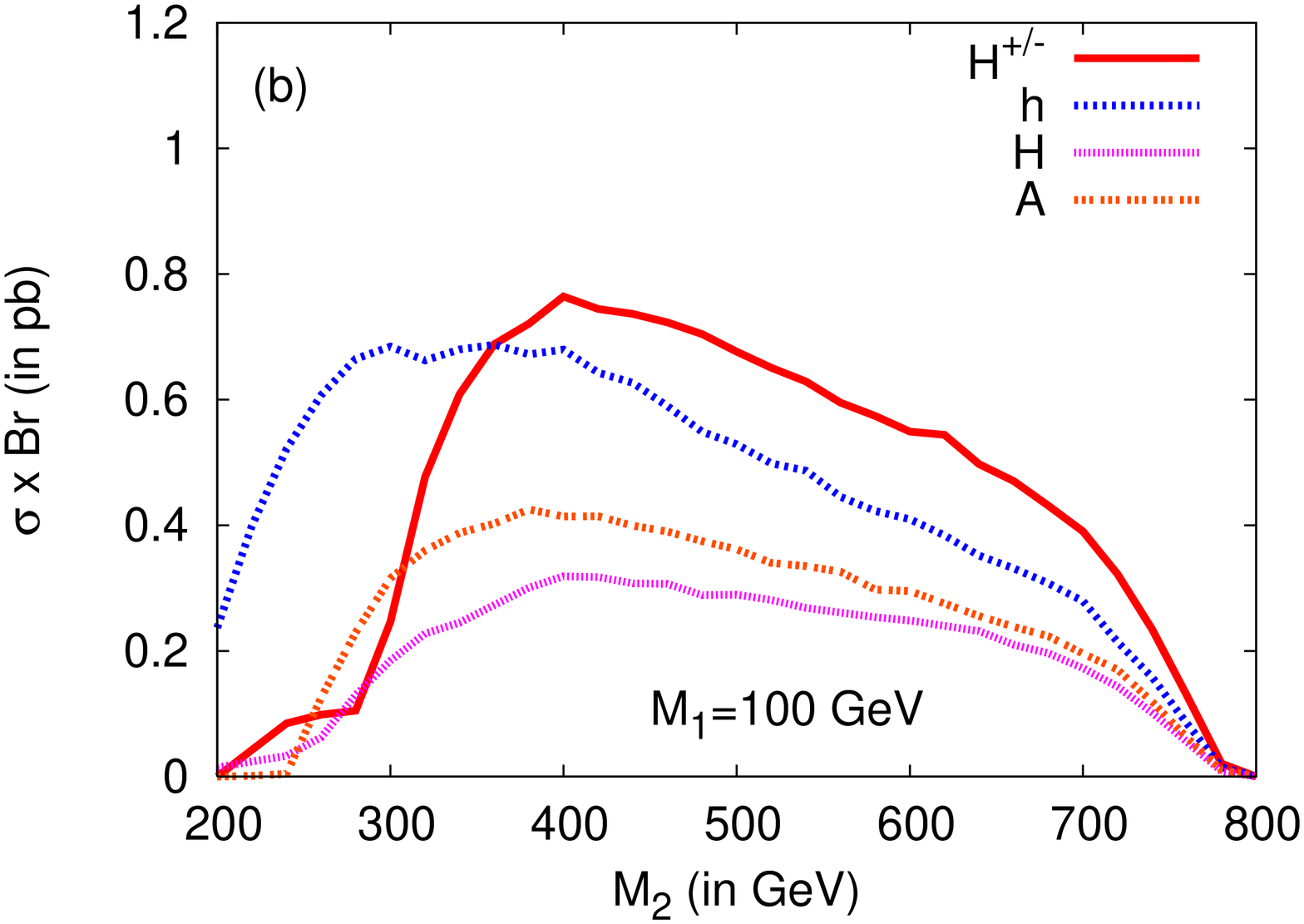,width=4.7cm,height=5.2cm,angle=-90.0}}
{\epsfig{file=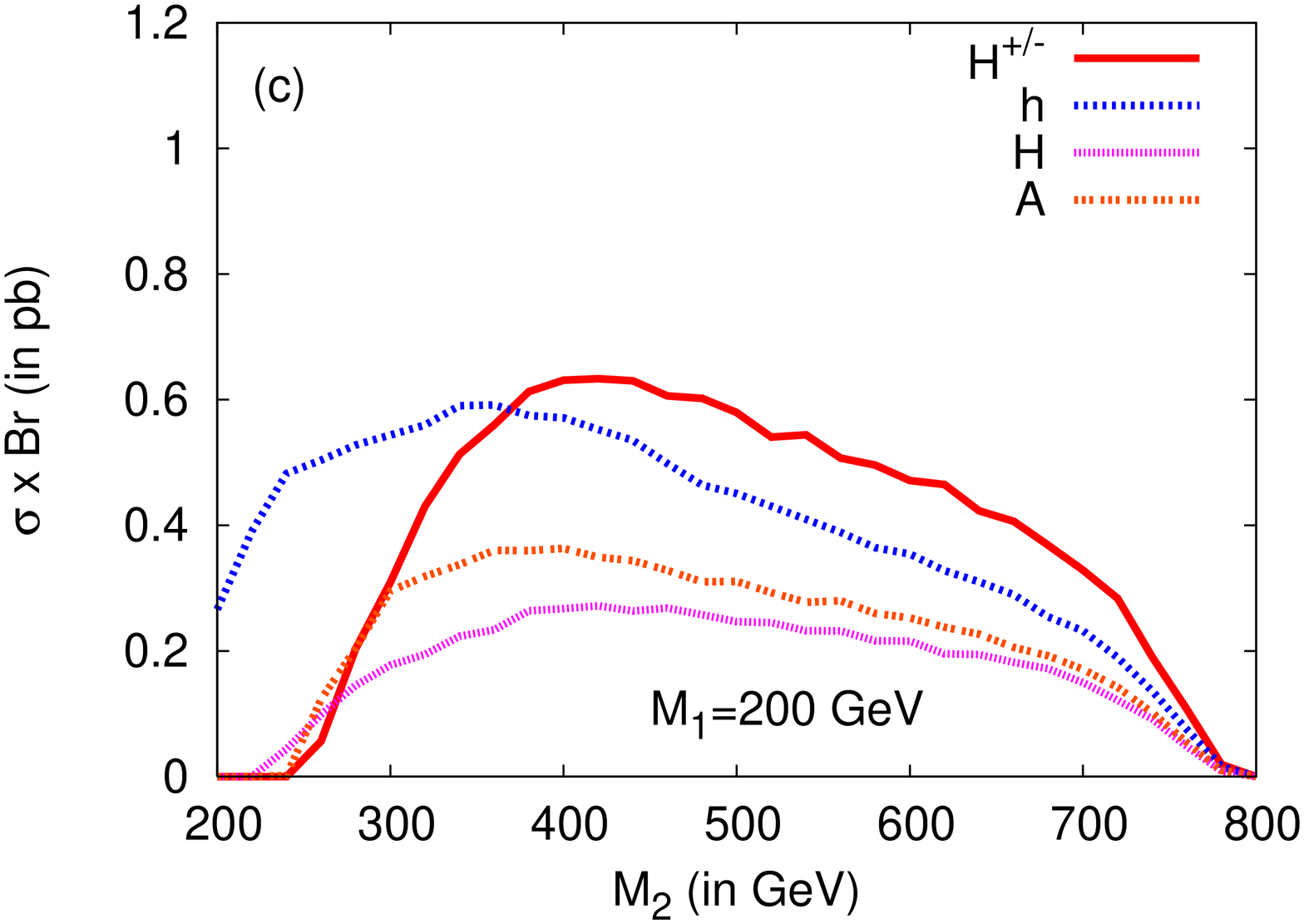,width=4.7cm,height=5.2cm,angle=-90.0}}
\caption{Effective cross-sections for universal ($\mone = \mtwo/2$, (a)) and 
nonuniversal (with $\mone=100$ GeV, (b) and with $\mone=200$ GeV (c)) for
$\mu=150$ GeV and $m_{H^{\pm}}=180$ GeV.} 
\end{center}
\label{fig11}
\vspace*{-1.0cm}
\end{figure}

\begin{figure}[hbt]
\begin{center}
{\epsfig{file=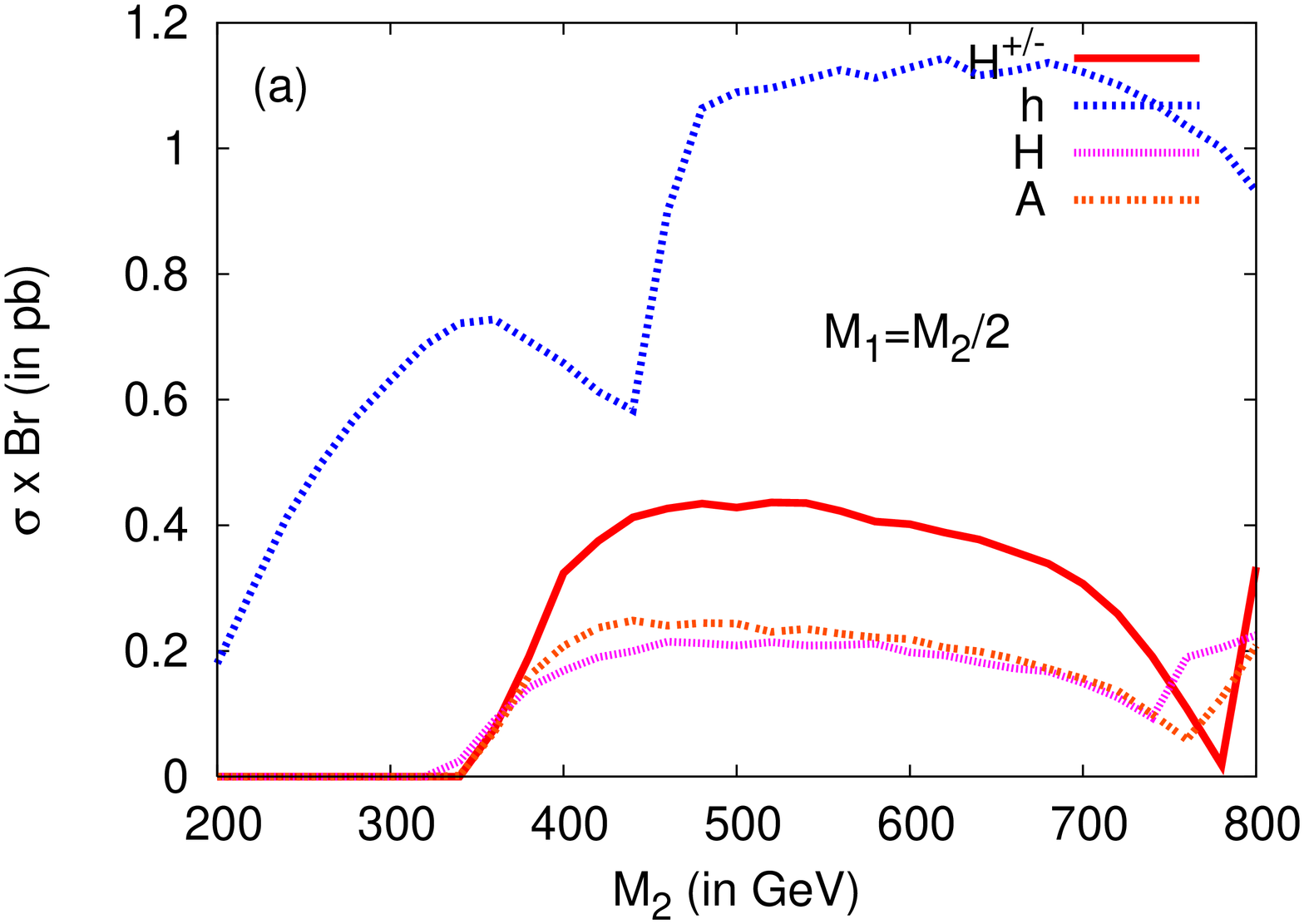,width=4.7 cm,height=5.2cm,angle=-90.0}}
{\epsfig{file=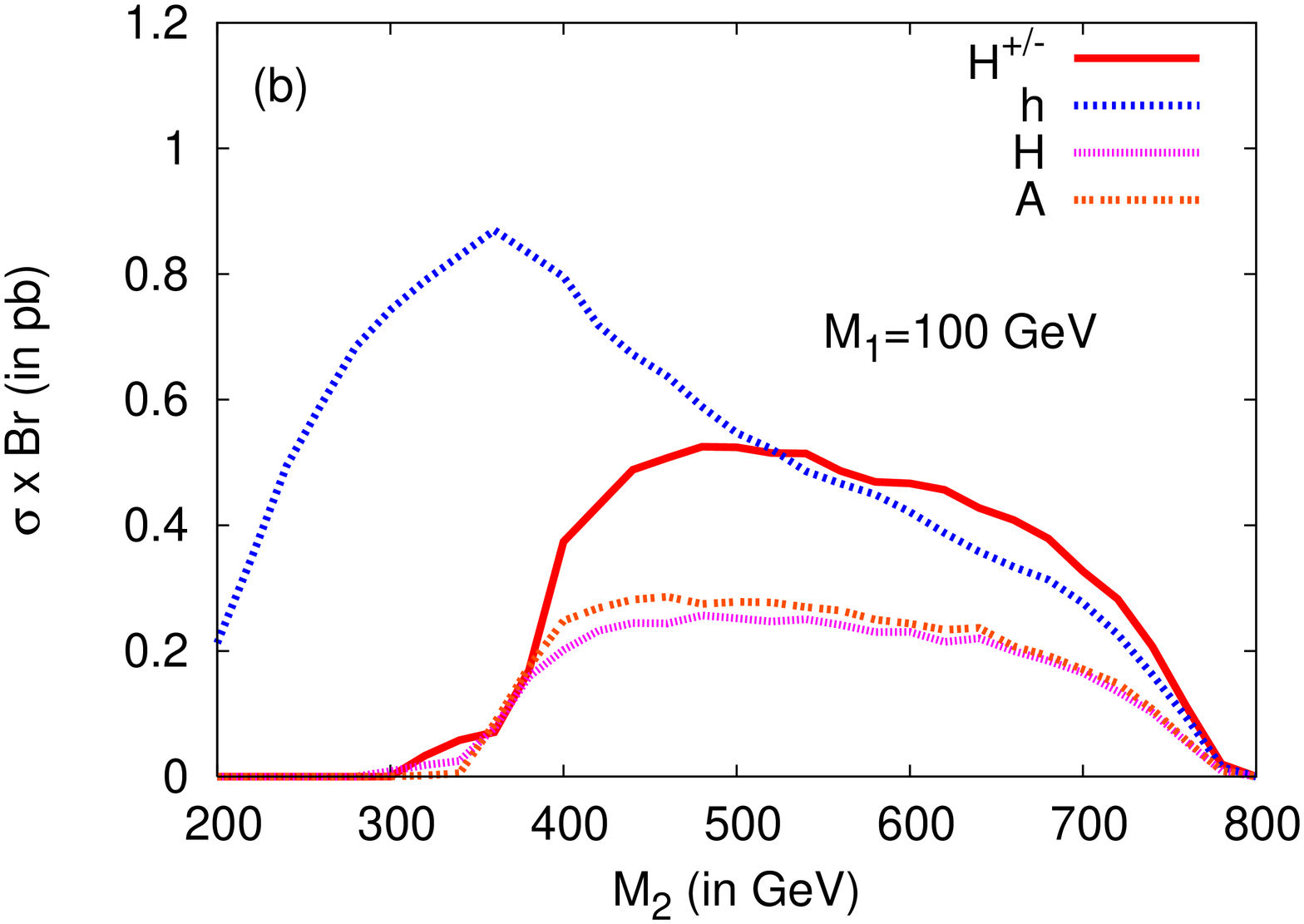,width=4.7cm,height=5.2cm,angle=-90.0}}
{\epsfig{file=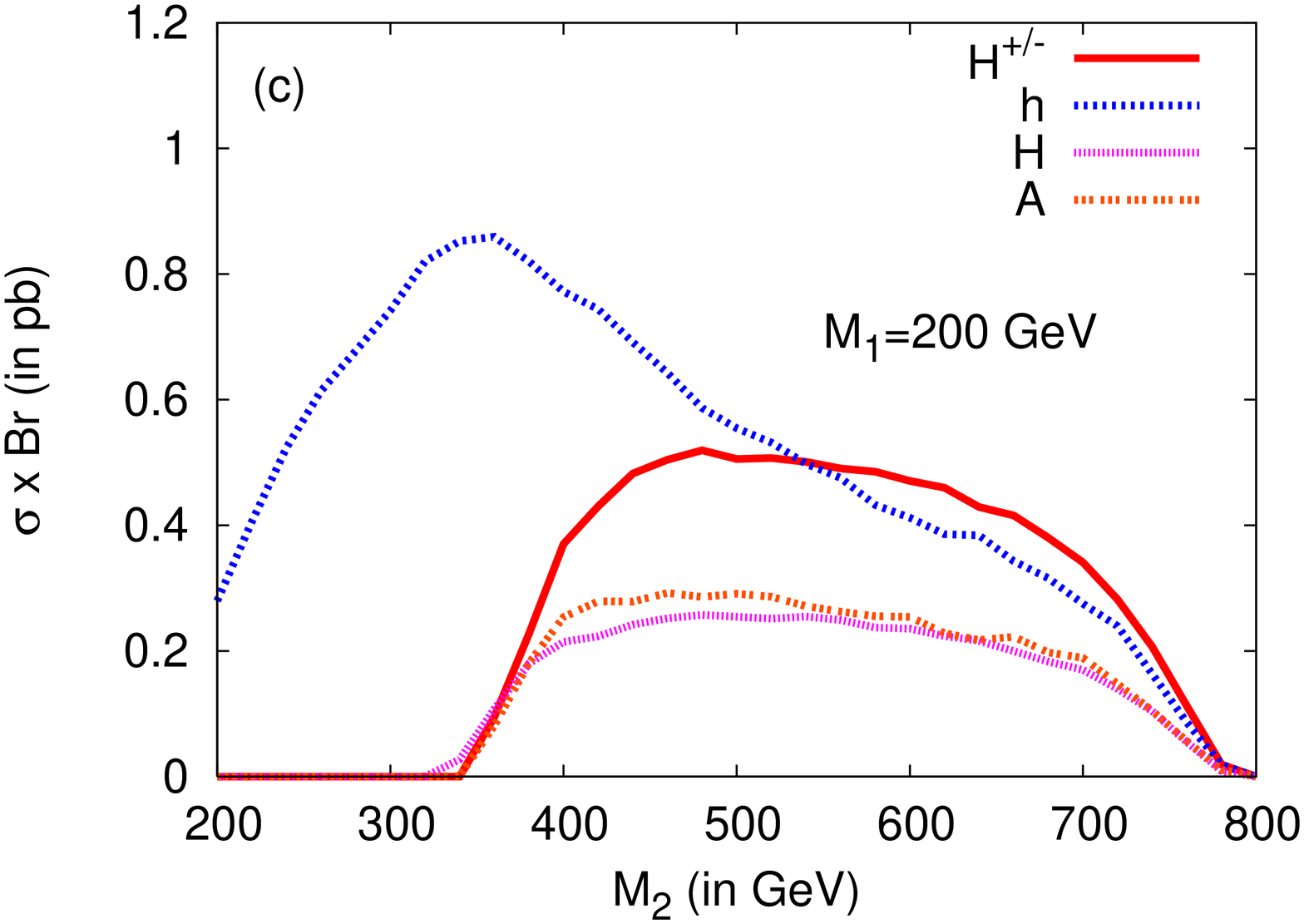,width=4.7cm,height=5.2cm,angle=-90.0}}
\caption{Effective cross-sections for universal ($\mone =
\mtwo/2$, (a)) and nonuniversal (with $\mone=100$ GeV, (b), with
$\mone=200$ GeV (c)) for $\mu=150$ GeV and $m_{H^{\pm}}=250$ GeV.} 
\end{center}
\label{fig11}
\end{figure}

\vspace*{0.8cm}
\begin{figure}[hbt]
\begin{center}
{\epsfig{file=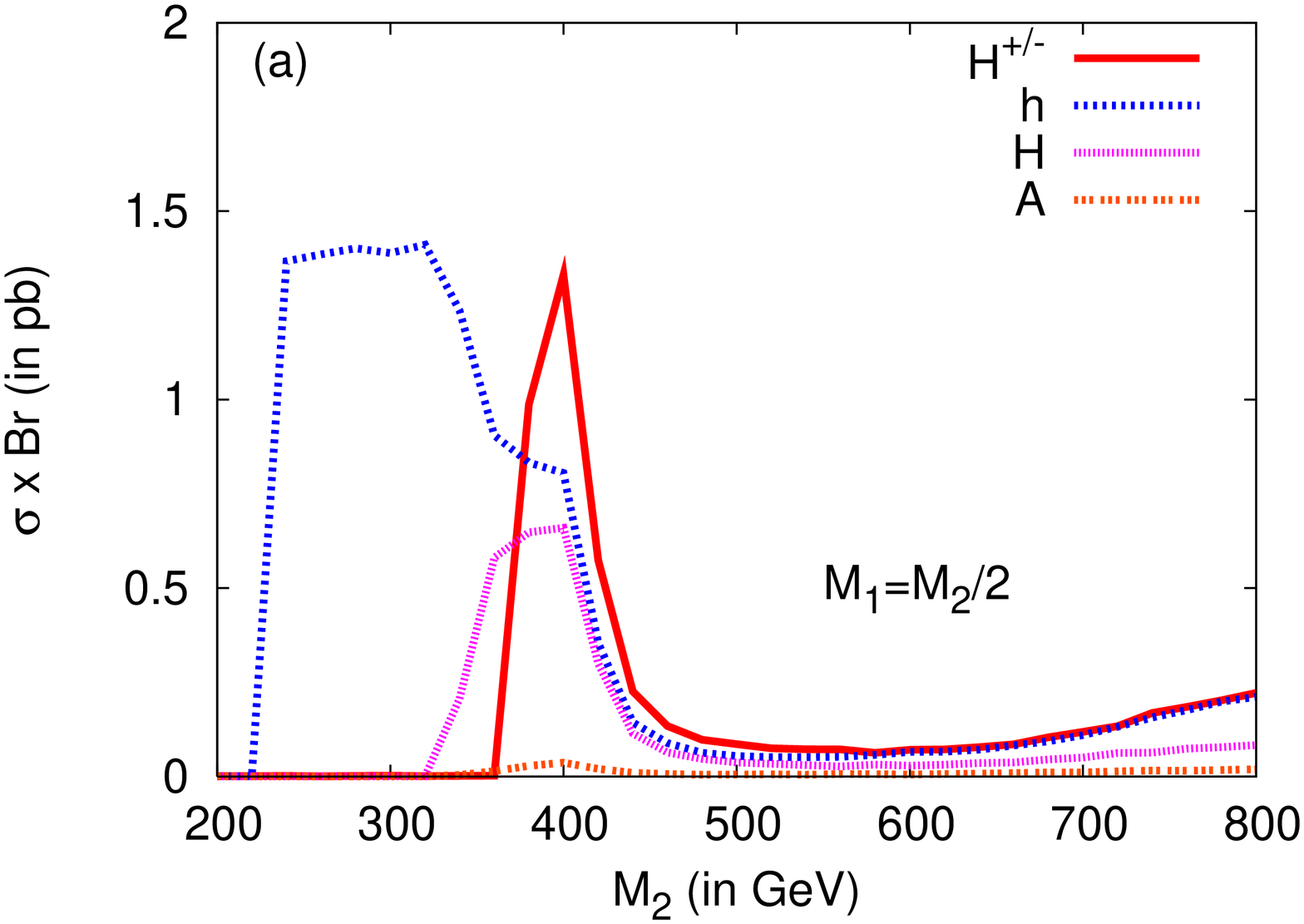,width=4.7 cm,height=5.2cm,angle=-90.0}}
{\epsfig{file=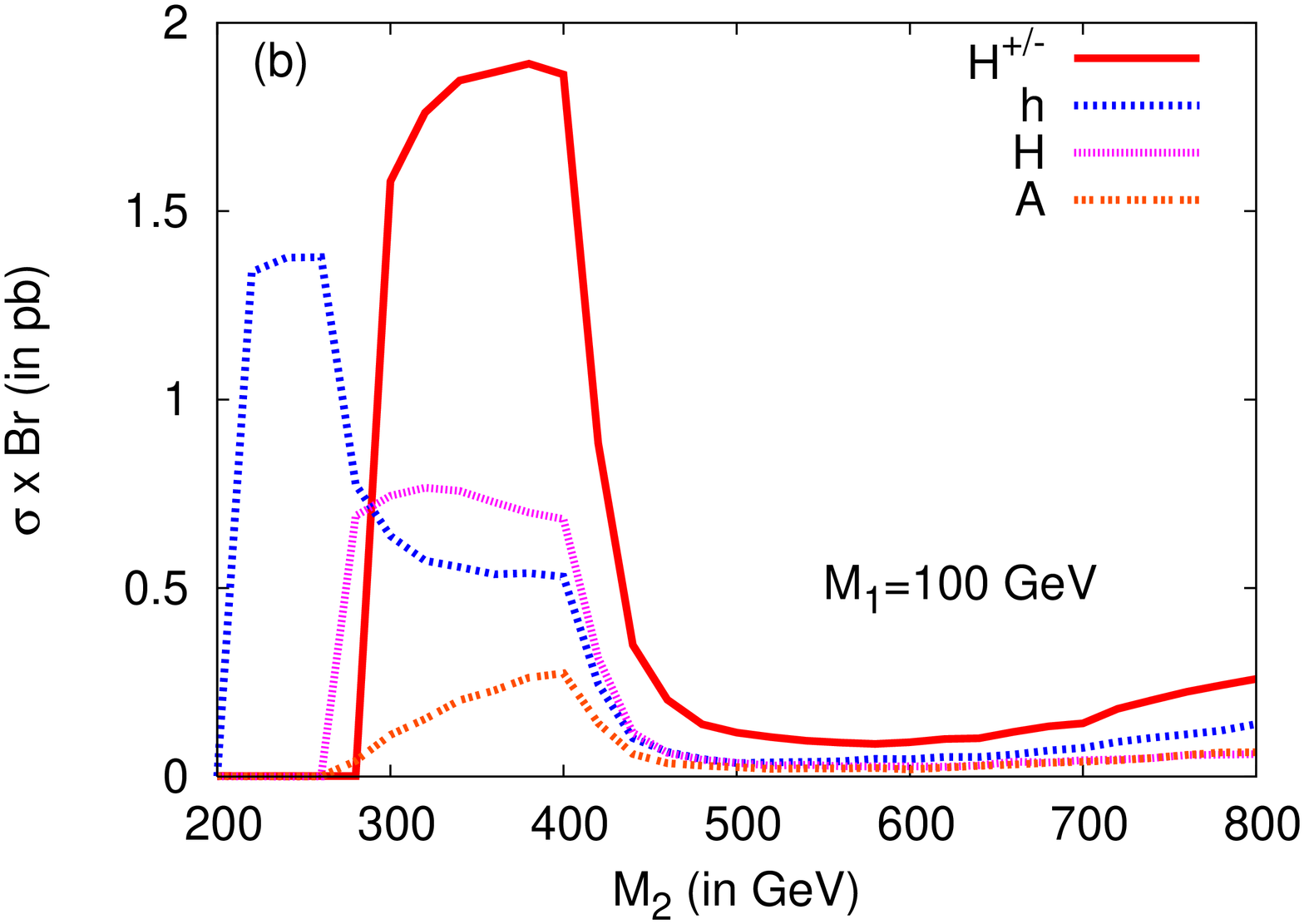,width=4.7cm,height=5.2cm,angle=-90.0}}
{\epsfig{file=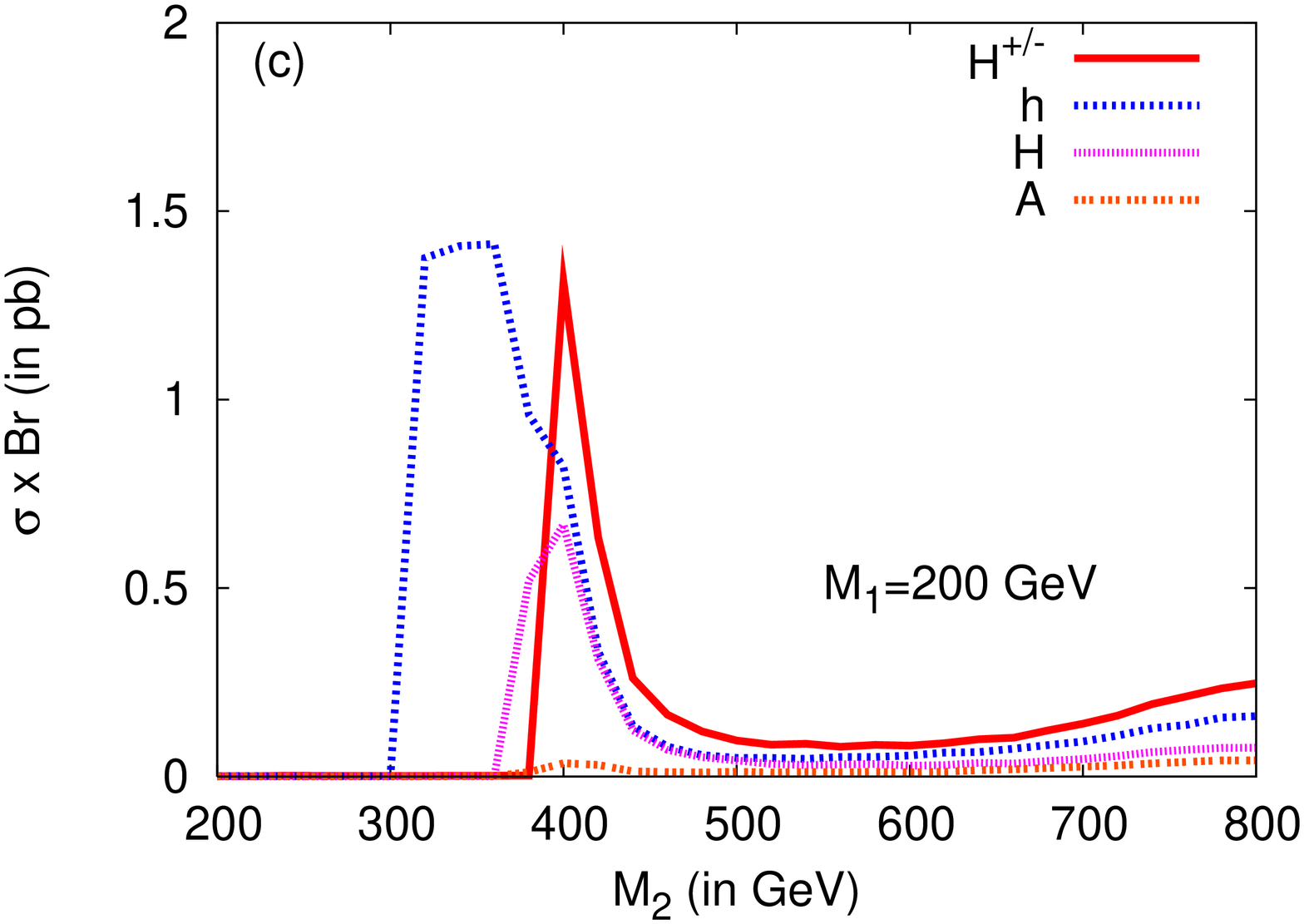,width=4.7cm,height=5.2cm,angle=-90.0}}
\caption{Effective cross-sections for universal ($\mone = \mtwo/2$,
(a)), nonuniversal (with $\mone=100$ GeV (b) and $\mone=200$ GeV (c)) for $\mu=700$ GeV and $m_{H^{\pm}}$=180 GeV.} 
\end{center}
\label{fig11}
\vspace*{-0.5cm}
\end{figure}

\begin{figure}[hbt]
\vspace*{-1.2cm}
\begin{center}
{\epsfig{file=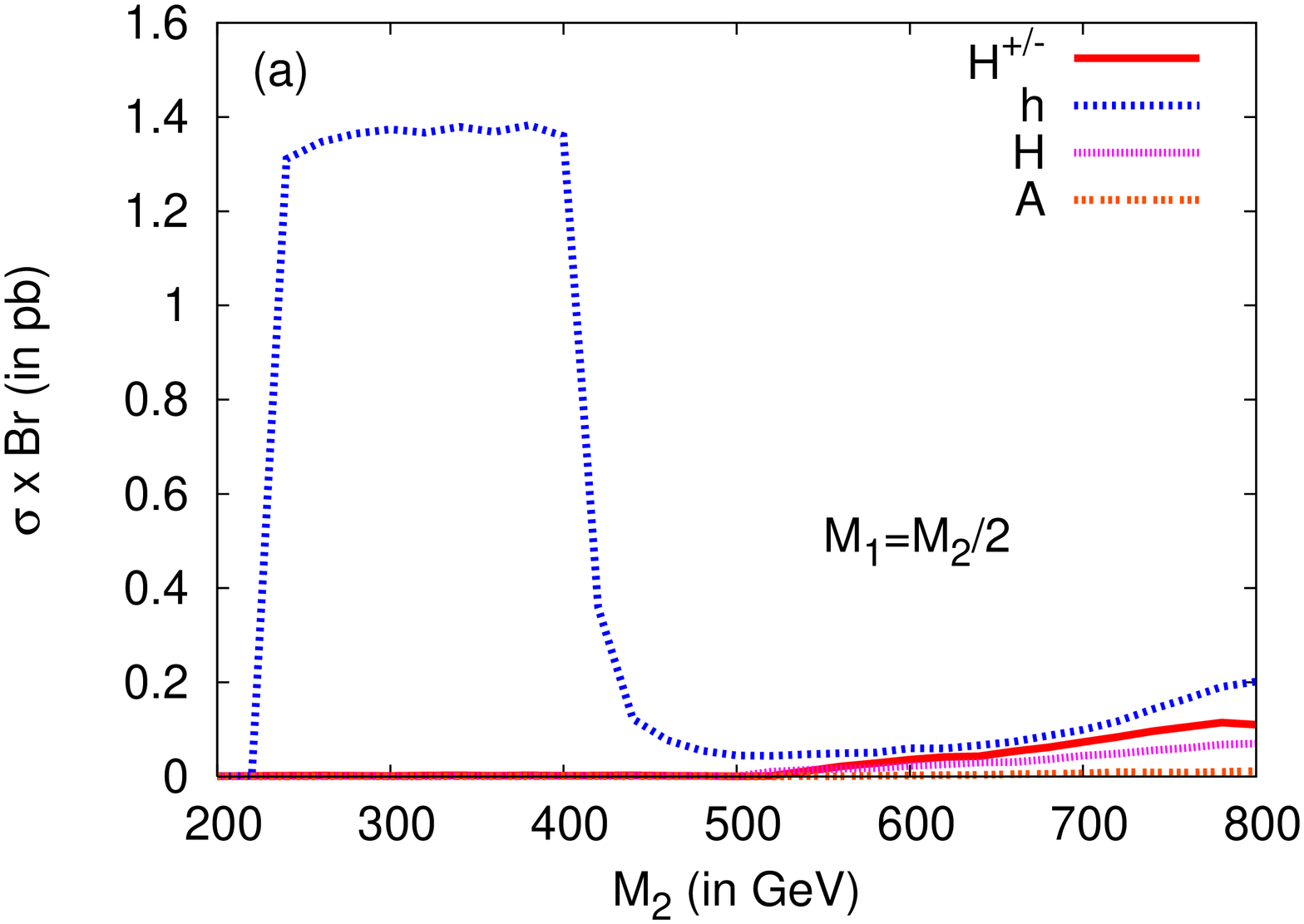,width=4.7 cm,height=5.2cm,angle=-90.0}}
{\epsfig{file=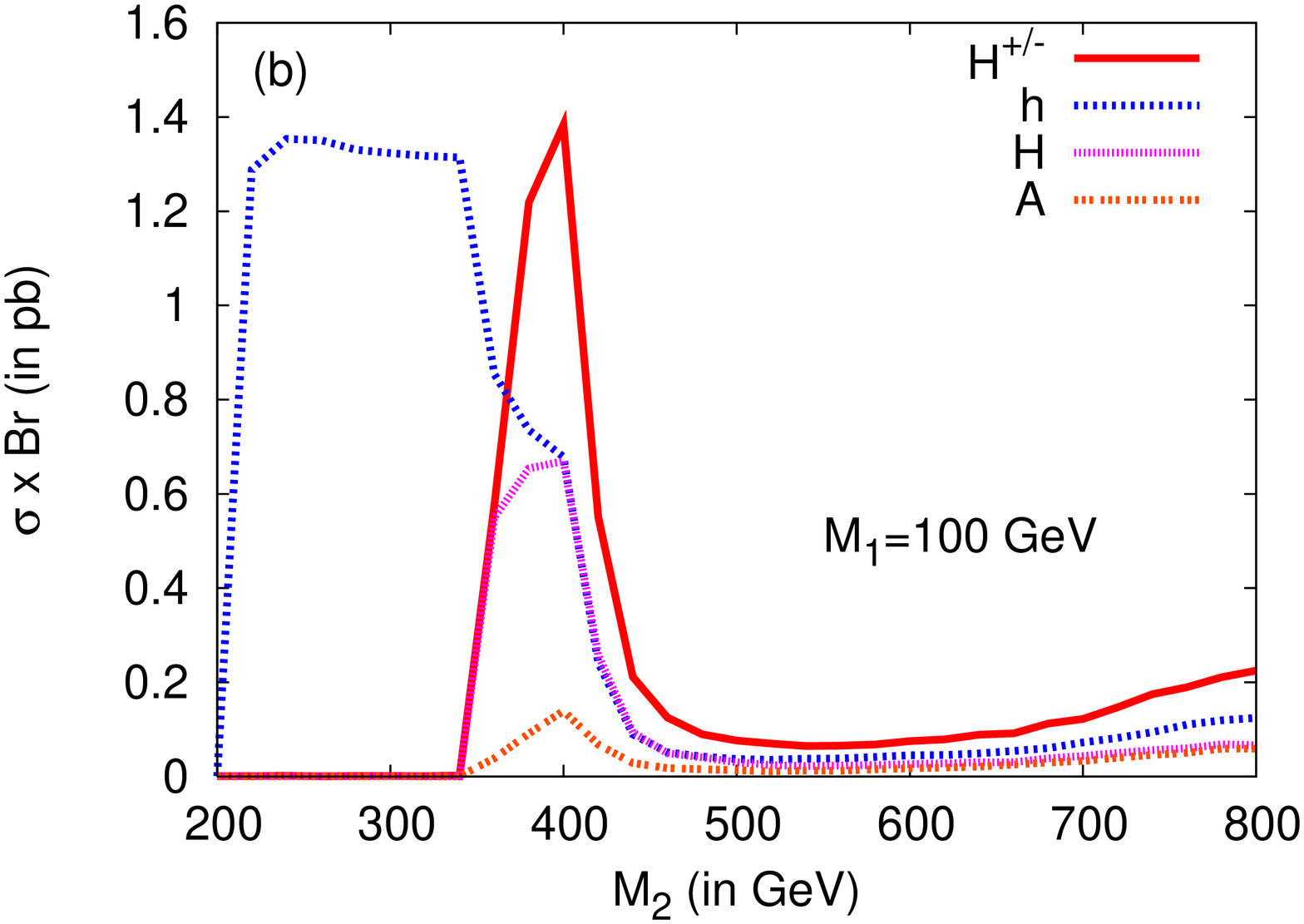,width=4.7cm,height=5.2cm,angle=-90.0}}
{\epsfig{file=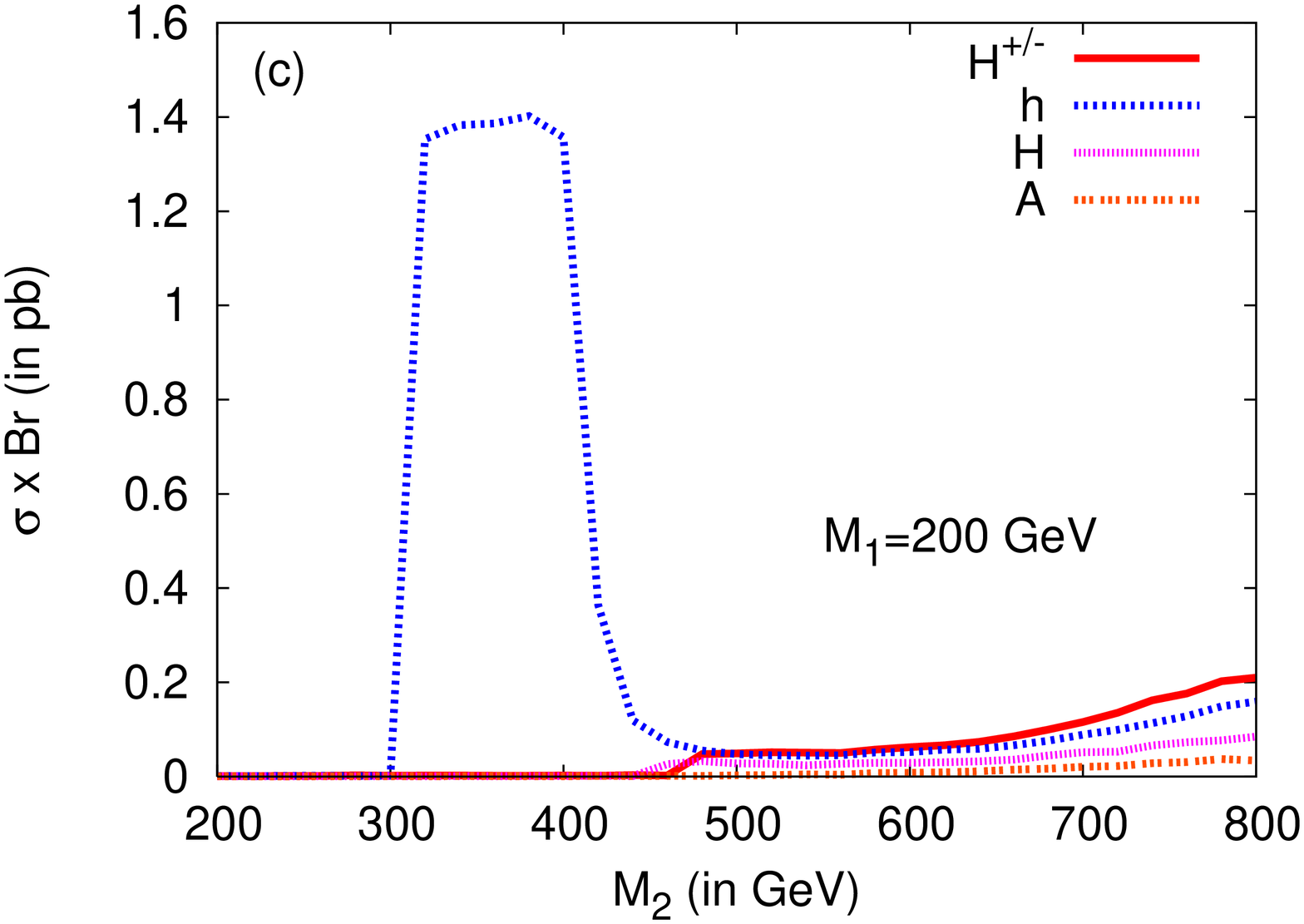,width=4.7cm,height=5.2cm,angle=-90.0}}
\caption{Effective cross-sections for universal ($\mone = \mtwo/2$,
(a)), nonuniversal (with $\mone=100$ GeV (b) and $\mone=200$ GeV (c))
 for $\mu=700$ GeV and $m_{H^{\pm}}$=250 GeV.} 
\end{center}
\label{fig11}
\hspace*{-0.5cm}
\end{figure}

Since $\mslep$ is taken to be 400 GeV, the
sleptons have a significant role in the cascades.
Thus, the natural expectation is that once these sleptonic
decay modes of the charginos/neutralinos open up, cascades to Higgs would get
suppressed. Hence the overall rates presented are of a conservative nature.
This is clear from the set of Figures 1 to 4. We discuss below some
generic features present in these figures and the information we get
from them.

Plot (a)s depict the universal scenario while plot (b)s and plot (c)s
are for the nonuniversal scenario. Figure 1 and Figure 2 are for $\mu=150$ GeV
while Figure 3 and Figure 4 are for $\mu=700$ GeV. For
$\mu=150$ GeV  the lighter gauginos are too closely degenerate
 (for both universal and nonuniversal scenarios) for the `little cascades'
to open up. Hence, the entire cascade contribution to Higgs production
comes from the `big cascade'. The sudden rises in some curves at
specific $\mtwo$ values in the universal scenario indicate attaining
the right mass-splitting between $\ntrl3$ and the LSP such that
$\ntrl3$ decaying to the lightest neutral (charged) Higgs boson and
the LSP (the lighter chargino) becomes possible. This feature is not
there in the nonuniversal cases (where $\mone$ is set to 100 and 200
GeV respectively) as $\mntrl3\sim \mu$.

One should note the different pattern of $M_2$-dependence of the
rates for $\mu=150$ GeV (Figures 1 \& 2) and 700 GeV (Figures 3 \& 4), respectively. This is because the
former situation allows Higgs production mostly through `big
cascades'. The latter case, where larger separation amongst the low-lying
states is possible, `little cascade' more abundantly, thus
making the variation of Higgs production rates with $M_2$ look
different. In particular, slepton masses of the order of 400 GeV affect
`big cascades' less for larger value of $M_2$ throught the enhancement
of the effective coupling for Higgs production. Little cascades are
affected much more in such a case, thus causing difference in the way
the rates fall with increasing $M_2$.

 In Figure 2 we illustrate a case similar to Figure 1 except for
$\mhpm=250$ GeV. This needs a substantial increment in the mass of
$A$ as input and results in a larger $H$ compared to those for Figure
1. Thus, in the universal case (Figure 2(a)) $\ntrl3$ needs to be heavier such that the heavier
 Higgs bosons may be produced in the decay of $\ntrl3$ along with
$\ntrl1$ or $\chpm1$. Note that increasing $\mhpm$ does not affect the
rate for the lightest higgs ($h$) significantly, when compared to the 
corresponding plot in Figure 1 since $m_h$ remains almost unaffected by
such an increase in $\mhpm$. Thus, for $\mhpm=250$ GeV, there is no
 cross-over between the curves for $h$ and $\hpm$ in the universal case
unlike $m_H^{\pm}$=180 GeV.

The general observation is that with increasing $\mhpm$ the threshold
value of $\mtwo$ shifts naturally to the right leading to more massive
heavier charginos and  neutralinos such that the `big cascades' may
take place. This eventually pushes the cross-over point (of the
rate-curves for the lightest and charged Higgs bosons) to  larger
values of $\mtwo$. Also, the rates for heavier Higgs bosons are
smaller for $\mhpm=250$ GeV as compared to $\mhpm=180$ GeV. There are
two reasons for this. First, the heavier Higgs bosons now become more
massive whose rates suffer a phase-space suppression for similar
chargino and neutralino masses. Second, the charginos and neutralinos
whose decay results in the Higgs bosons (with increased masses) have to
become heavier as well. Thus the production
rates for the latter also get affected. Figures 2(b) and 2(c) illustrating
the nonuniversal cases with $\mone=100$ GeV and 200 GeV are to be
contrasted with the corresponding ones in Figure 1. They only differ
by the generic features as described above.

Figure 4 illustrates a situation similar to Figure 3 except for $\mhpm=250$
GeV. In the universal case (Figure 4(a)), the peak in the $\hpm$ rate 
disappears when compared to Figure 3(a). One should note that this peak is due
predominantly to a `little cascade' like $\chpm1 \to \ntrl1 \hpm$.
As indicated earlier, with growing $m_{\hpm}$ the mass splitting between 
the gaugino states involved above are not enough to accommodate the
above cascade. The situation could be a little different for a
nonuniversal case with a lower value of $\mone$ (100 GeV) as shown in
 Figure 4(b). The smaller value of $\mone$ now ensures a lower mass for the
$\ntrl1$ thus help regaining the required splitting when the peak in 
the $\hpm$ is back (at around 400 GeV). In Figure 4(c), $\mone$
is 200 GeV and this again blocks the above decay mode at around 400
GeV. Of course, with increasing value of $\mtwo$ ($\geq 500$ GeV) the
Higgs productions under such cascades open up again. But this time,
 Br[$\chpm1 \to \ntrl1 \hpm$] starts getting suppressed as the two-body sleptonic decay modes
 of $\chpm1$ take off for our choice of slepton mass (400 GeV).

\subsection{Variation with $\mu$}

\begin{figure}[hbt]
\begin{center}
\vspace*{-0.5cm}
\hskip -65pt
{\epsfig{file=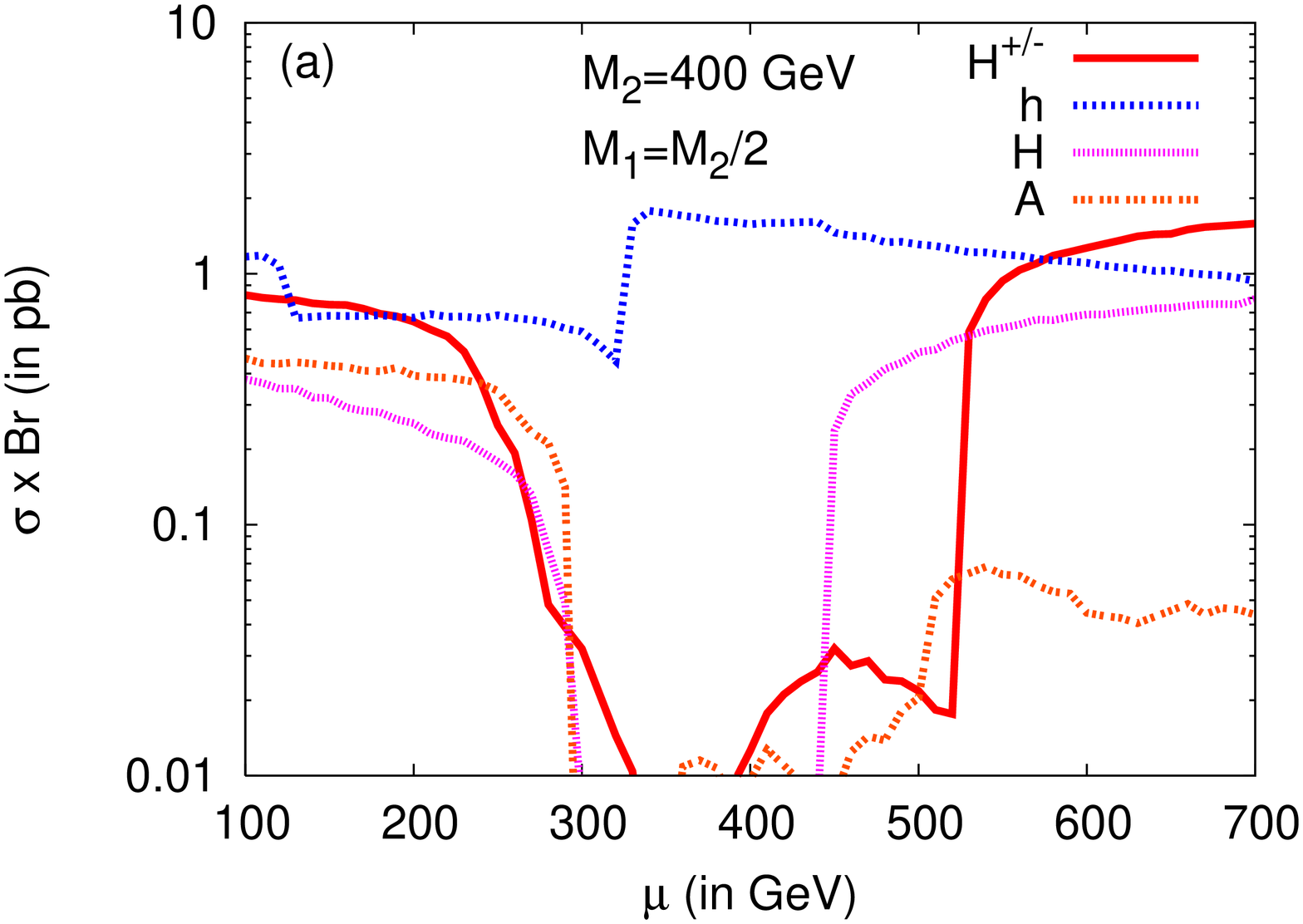,width=6.0 cm,height=6.7cm,angle=-90.0}}
{\epsfig{file=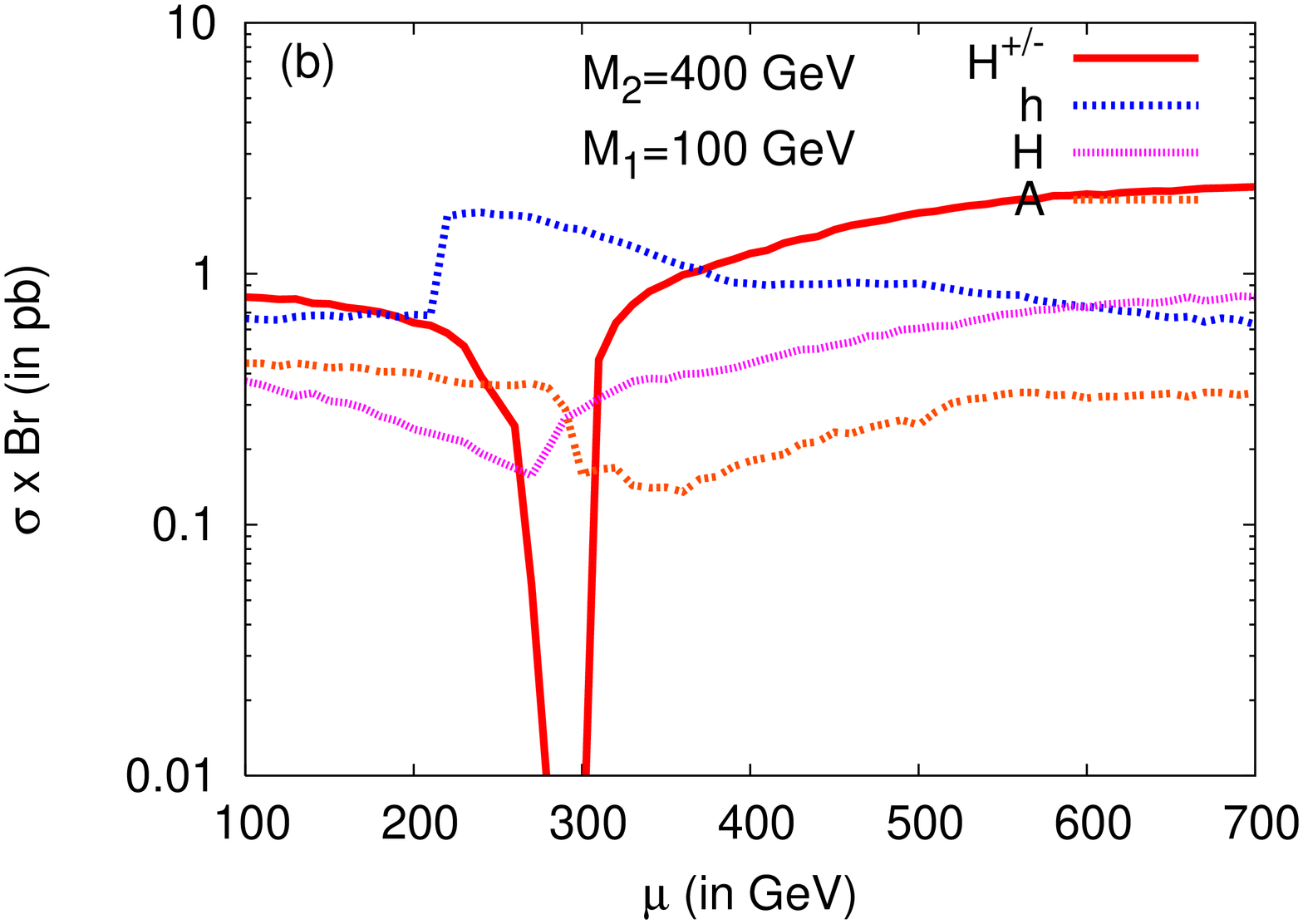,width=6.0 cm,height=6.7cm,angle=-90.0}}
\caption{Effective cross-section as a function of $\mu$ for universal
(with $\mone=\mtwo/2$, (a))  and nonuniversal (with $\mone=100$ GeV, (b)) 
scenarios with  $\mtwo=400$ GeV and $m_{H^{\pm}}=180$ GeV.} 
\end{center}
\label{fig11}
\end{figure}

\begin{figure}[hbt]
\begin{center}
\vspace*{-0.5cm}
\hskip -65pt
{\epsfig{file=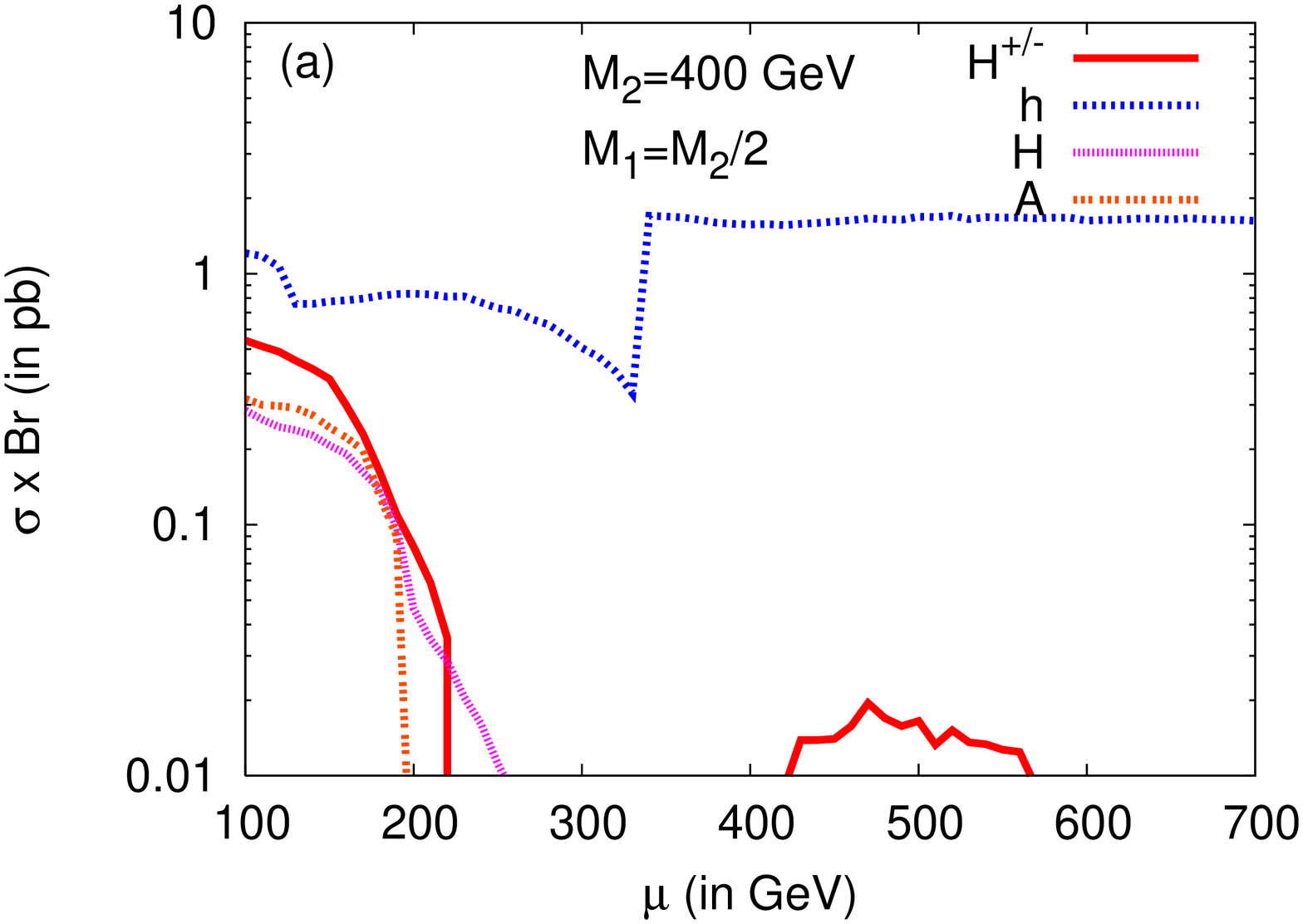,width=6.0 cm,height=6.7cm,angle=-90.0}}
{\epsfig{file=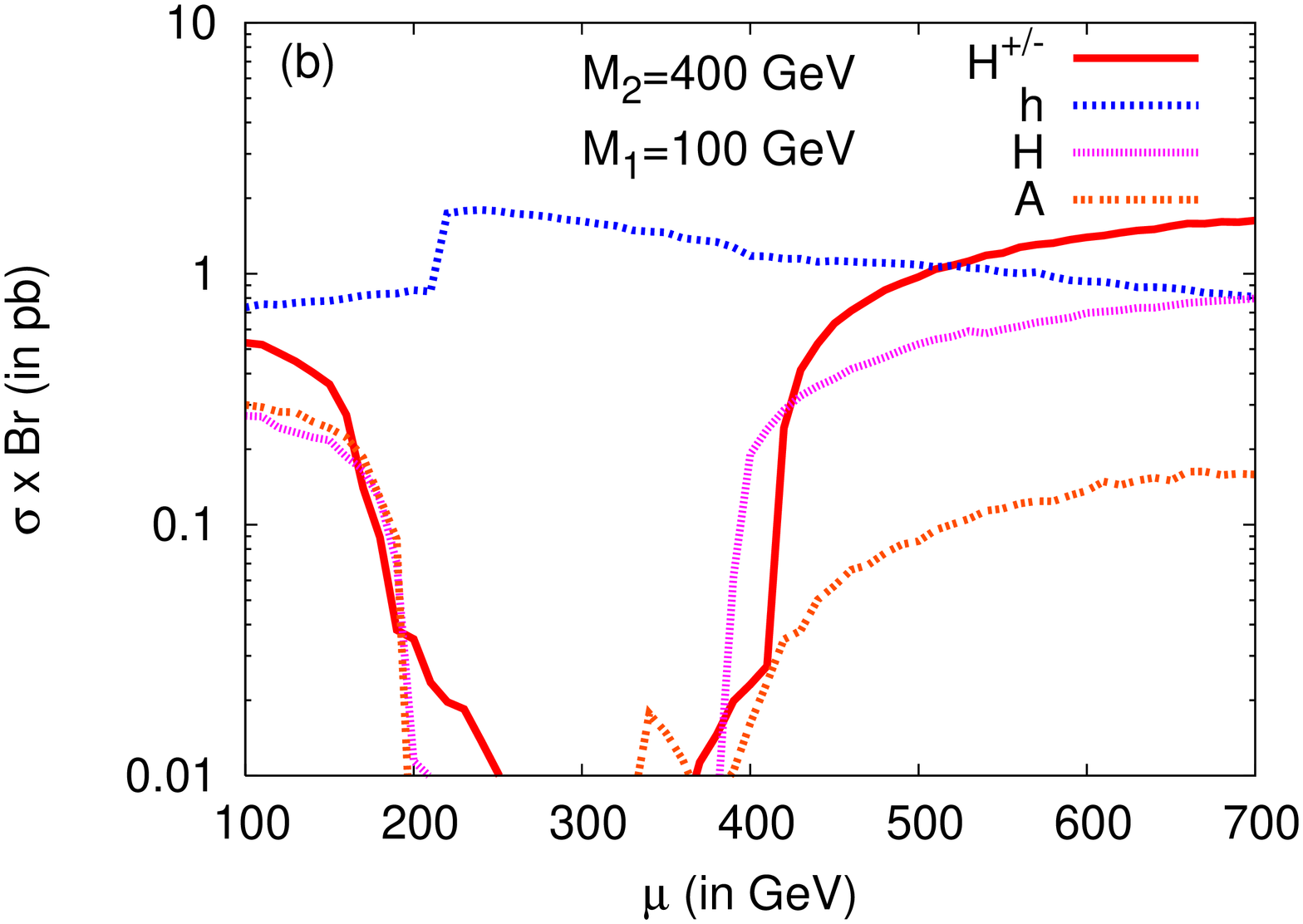,width=6.0cm,height=6.7cm,angle=-90.0}}
\caption{Effective cross-section as a function of $\mu$ for universal
(with $\mone=\mtwo/2$, (a))  and nonuniversal (with $\mone=100$ GeV, (b)) 
scenarios with  $\mtwo=400$ GeV and $m_{H^{\pm}}=250$ GeV.} 
\end{center}
\label{fig11}
\vspace*{-1.0cm}
\end{figure}

In Figures 5 \& 6, we illustrate the variation of the rates for Higgs production under 
SUSY cascades with $\mu$. Plot (a)s represent the universal scenario
while plot (b)s illustrate the same in a nonuniversal situation. In both
cases, for small values of $\mu$, the lighter charginos and
neutralinos are  Higgsino-like and their masses are of the order of
$\sim \mu$ with a definite split from their heavier mates governed by
the value of $M_2$ chosen. With increasing $\mu$, the ``mixed region'' is approached
and the mass-splittings among the heavier and lighter charginos
decrease. This gradually eliminates  the possible sources of the
heavier Higgs bosons, especially those of $\hpm$. Further, with
increasing $\mu$, channels like $\chpm2 \to \ntrl1 \hpm$ again opens
up followed by $\chpm1 \to \ntrl1 \hpm$ at about $\mu=525$ GeV in the 
universal case. A further mild rise in the $\hpm$-rate is observed at
 around $\mu=550$ GeV when $\ntrl3 \to \chpm1 \hpm$ opens up. However,
in the nonuniversal case, due to a fixed low value of $\mone$ the LSP mass 
remains almost the same ($\sim \mone$). Hence channels like $\chpm1 \to
\ntrl1 \hpm$ opens up as early as $\mu=300$ GeV. The features observed
in  these plots are generic as long as the relative splittings of
gaugino masses and $\mu$ used here remain similar.

 In Figure 6 we illustrate the case similar to Figure 5 except for 
$\mhpm=250$ GeV. In the universal case, when compared to Figure 5, we find
that the rise in the rate for $\hpm$ at around $\mu=500$ GeV is missing.
This is because of the lack of enough splitting between the lighter
chargino and the LSP masses that might lead to $\hpm$ unlike what happened in 
the universal case when $\mhpm=180$ GeV. For $\mone=100$ GeV, $\chpm1
\to \ntrl1 \hpm$ opens up at a later stage compared to $m_H^{\pm}$=180
GeV case. For higher neutral Higgses similar things happen.

\subsection{Comparison between $\hpm$ and $h$ production rates} 
Figures 1, 3 and 5 ($m_{H^{\pm}}$=180 GeV) along with the discussions
in sections 3.1 and 3.2 reveal a rather complicated interplay of
several masses and couplings. It is undoubtedly a difficult task
 to extract useful information from  these processes in a systematic
way. However, a closer look at these reveal that we have a somewhat
 clean feature in the variations of rates for $\hpm$ and $h$ which we
 can use to our benefit. These rates behave in a distinctly
complementary fashion as functions of $\mtwo$ (Figures 1 and 3) and
 $\mu$ (Figure 5) and show up as multiple cross-over points. For
universal and nonuniversal scenarios, these 
cross-overs take place at different values of $\mtwo$ and $\mu$. This can be
understood in a straight-forward manner following the discussions in sections 3.1 and 3.2.

The observations prompt us to look out for a suitable illustration of
the situation. In Figure 7 we present the scattered plots of the
two-rates compared in the $\mtwo-\mu$ plane for both universal and 
nonuniversal cases. Over the grey regions the rates for the charged
Higgs bosons is greater than that for the lightest neutral Higgs boson and
the reverse is the case over the darker regions. At suitable points on
these plots, direct correspondences can be made with Figures 1, 3 and
5. There are quite a few distinctive patches on these two plots. If
such relative rates can be known from the LHC experiments, these
plots could be used to have a preliminary understanding of the gaugino
mass pattern.

 In Figure 8 we present a similar set of scattered plots but for
$m_H^{\pm}$=250 GeV. In the universal case, all over the $\mtwo-\mu$ space
the rate for lightest Higgs is greater than that for the charged Higgses.
For the nonuniversal case, with increasing $\mone$ the region where the 
rate for the charged Higgses is larger than that for the lightest Higgs 
gradually shrinks. Again, all these can be read off from the plots in Figures
1 to 6. Thus, we see that $\mhpm$ and $\mone$ play some significant roles
in characterizing the $\mtwo-\mu$ plane.

We have checked that the results presented here are robust against
wide variation of squark and gluino masses, even when their ordering
is reversed. Increasing the slepton masses would only make life
simpler in the sense that they get decoupled from the cascades and we
 checked that this does not bring out any new feature. We also
checked that the effect of $\tan\beta$ is anything but significant. 
This validates one of the very basic expectations for studying the
Higgs  production under SUSY cascades and hence its power 
of probing the Higgs sector for intermediate values of $\tan\beta$.
 
It is useful to mention here that the confirmation of gaugino
nonuniversality in Higgs production presupposes some knowledge of
the SUSY spectrum, as obtained from the heaviness or
missing-${E_T}$ distribution of the SUSY signals. As the scatter plots
in figures 7 and 8 indicate, one is able to infer conclusively about
 universality (or the lack of it) from the relative magnitudes of the
 $h$- and $H^{\pm}$-production rates, provided that two quantities out
 of $M_1$, $M_2$ and $\mu$ are known. However, even that does not rob
the present study of its merits, since any unambiguous conclusion
about the particle spectrum of a new physics scenario requires a
multichannel analysis. In that sense, the inclusive Higgs production
 processes provide a channel of supreme importance in understanding the
electroweak gaugino-Higgsino sector.

\vspace*{0.5cm}
\begin{figure}[hbt]
\begin{center}
\vspace*{-0.5cm}
\hskip 0pt
{\epsfig{file=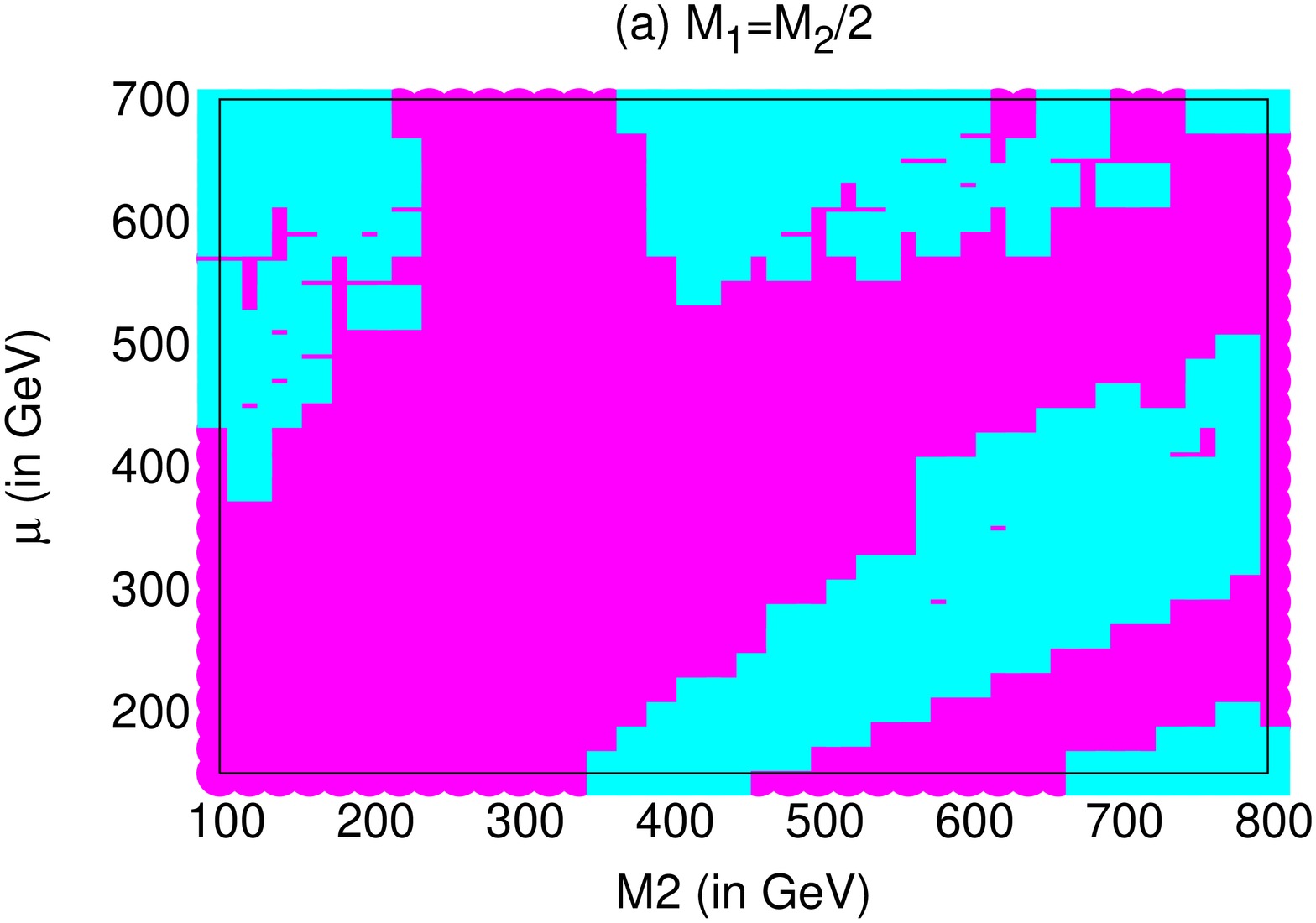,width=5.0 cm,height=5.2cm,angle=-90.0}}
\hskip 2pt
{\epsfig{file=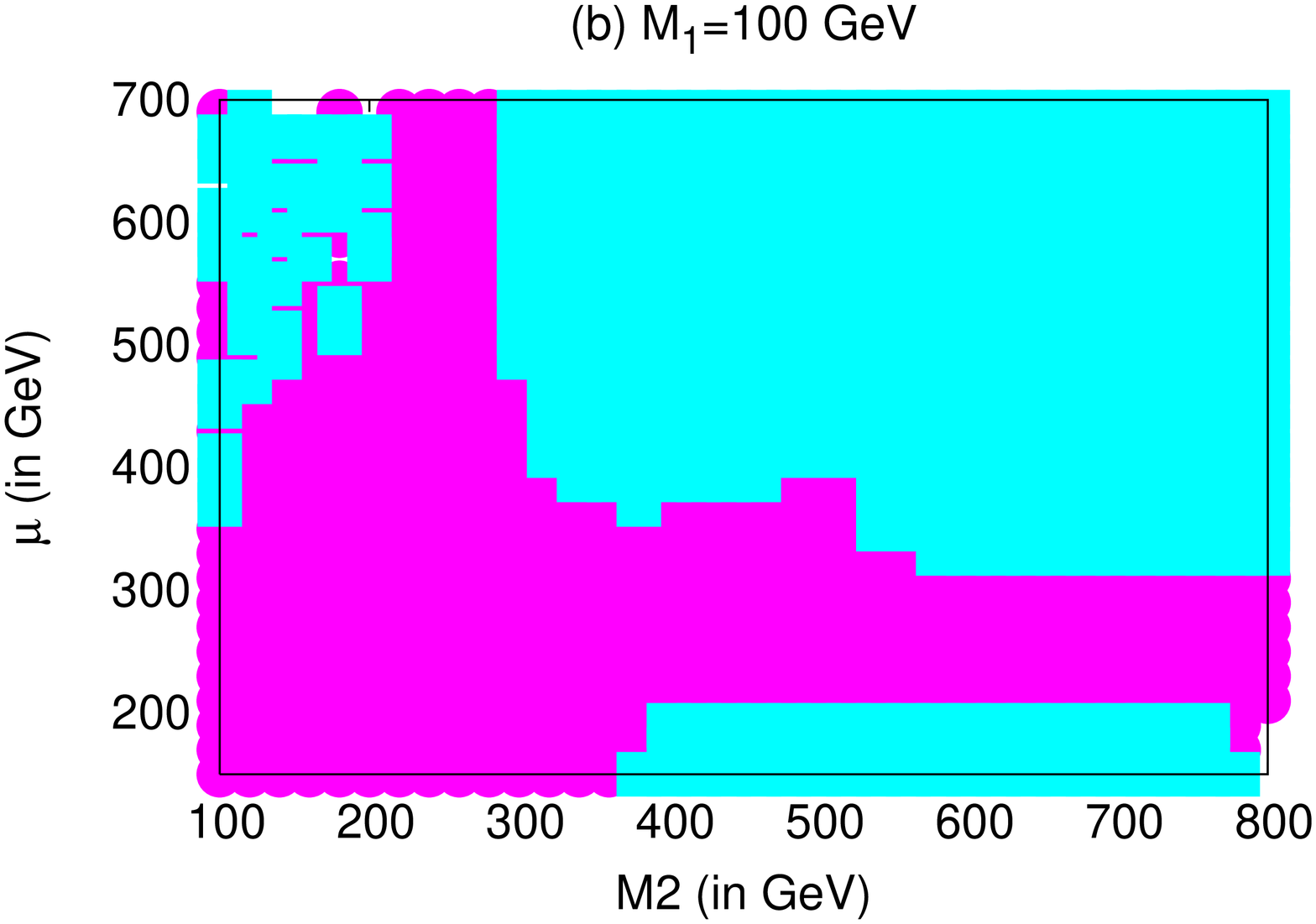,width=5.0 cm,height=5.2cm,angle=-90.0}}
\hskip 2pt
{\epsfig{file=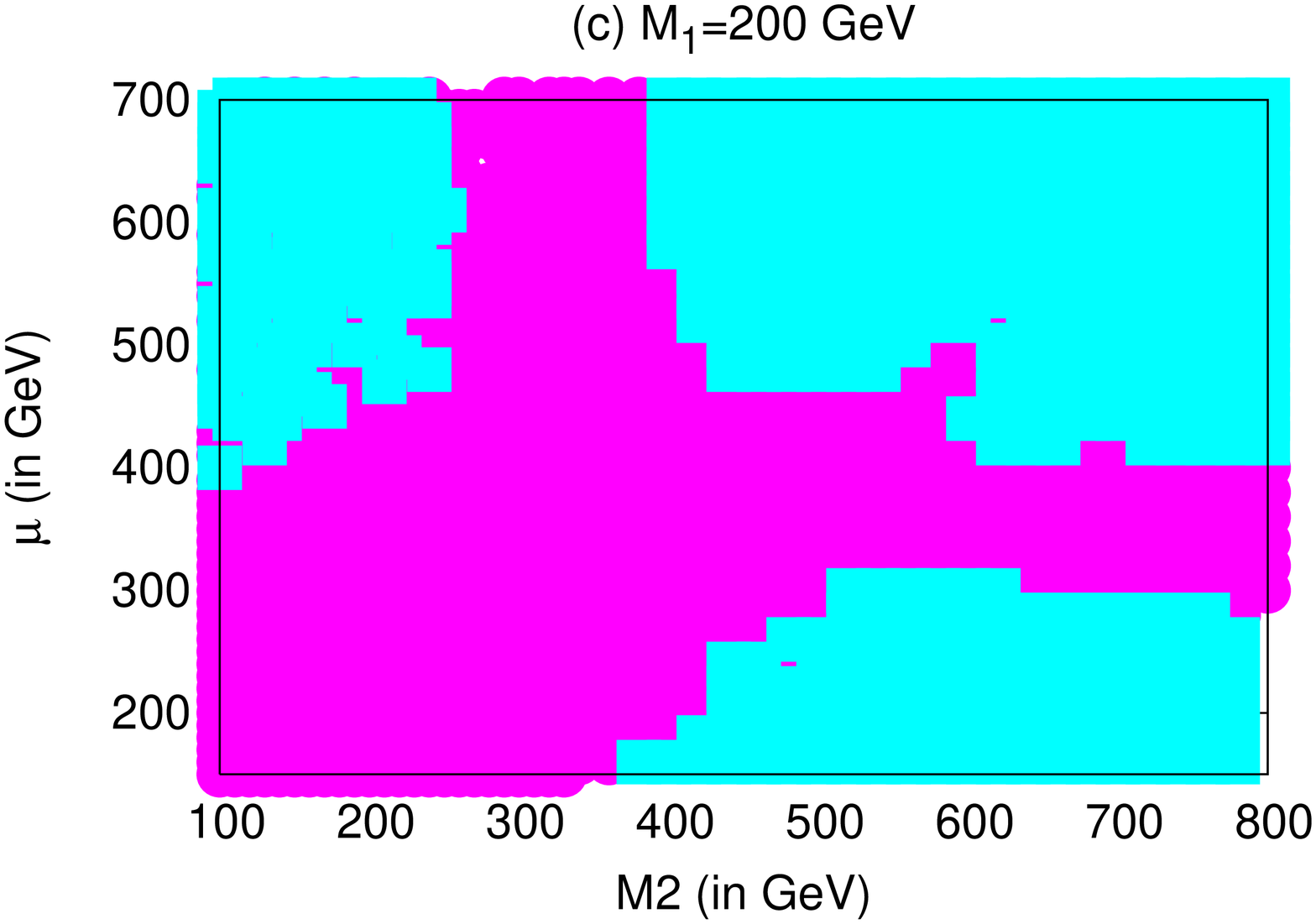,width=5.0 cm,height=5.2cm,angle=-90.0}}	
\caption{The pink (dark) region is the region where the rate for H$^{\pm}$ 
is less than that for $h$ production, for the sky-blue region (grey) 
it is the opposite and white regions indicate zero production for both.
We take $\mone=\mtwo/2$ (a), $\mone=100$ GeV (b) and $\mone=200$ GeV
(c). $m_{H^{\pm}}$ is set to 180 GeV.} 
\end{center}
\label{fig11}
\end{figure}

\begin{figure}[hbt]
\begin{center}
\vspace*{-0.5cm}
\hskip 0pt
{\epsfig{file=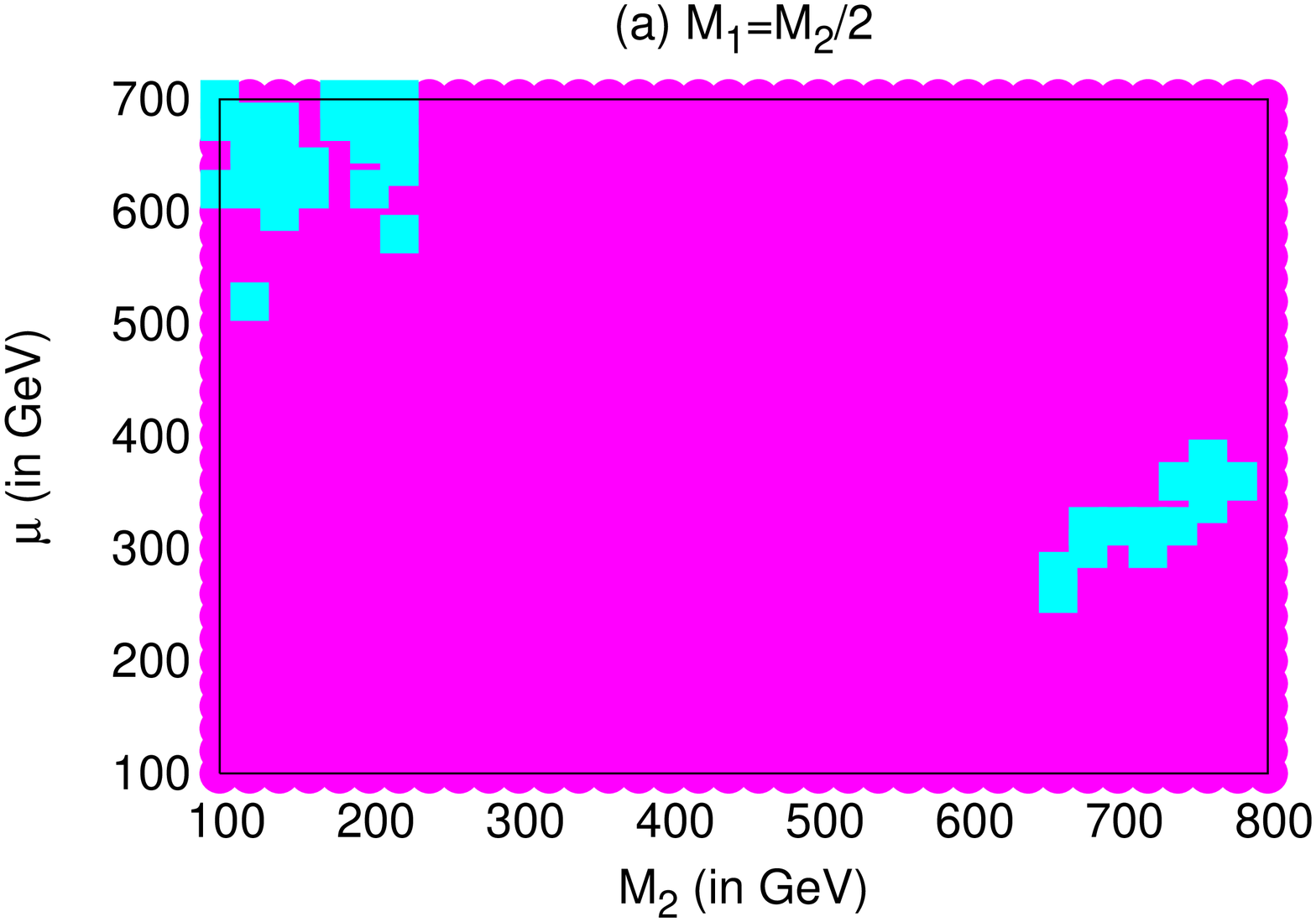,width=5.0 cm,height=5.2cm,angle=-90.0}}
\hskip 2pt
{\epsfig{file=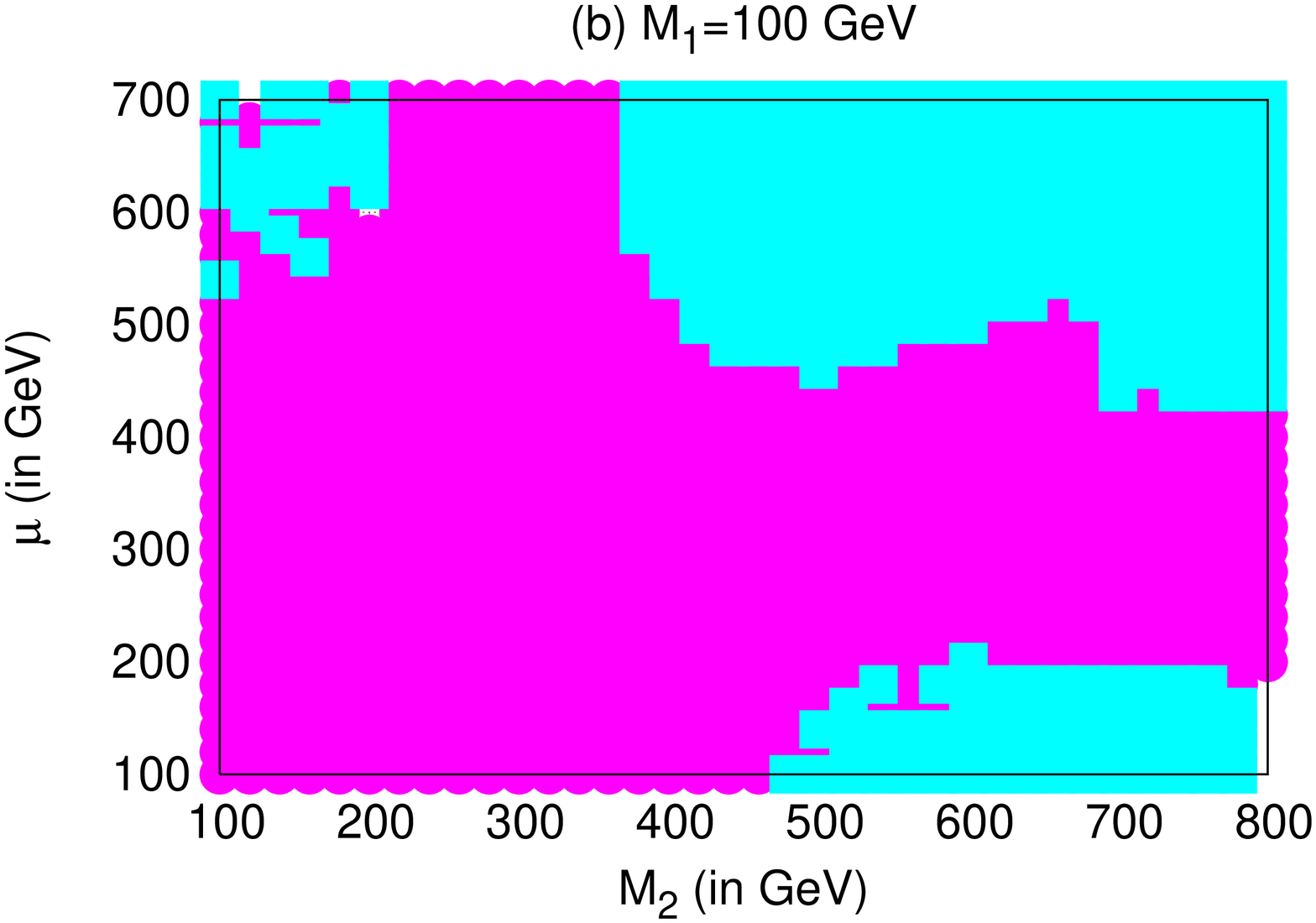,width=5.0 cm,height=5.2cm,angle=-90.0}}
\hskip 2pt
{\epsfig{file=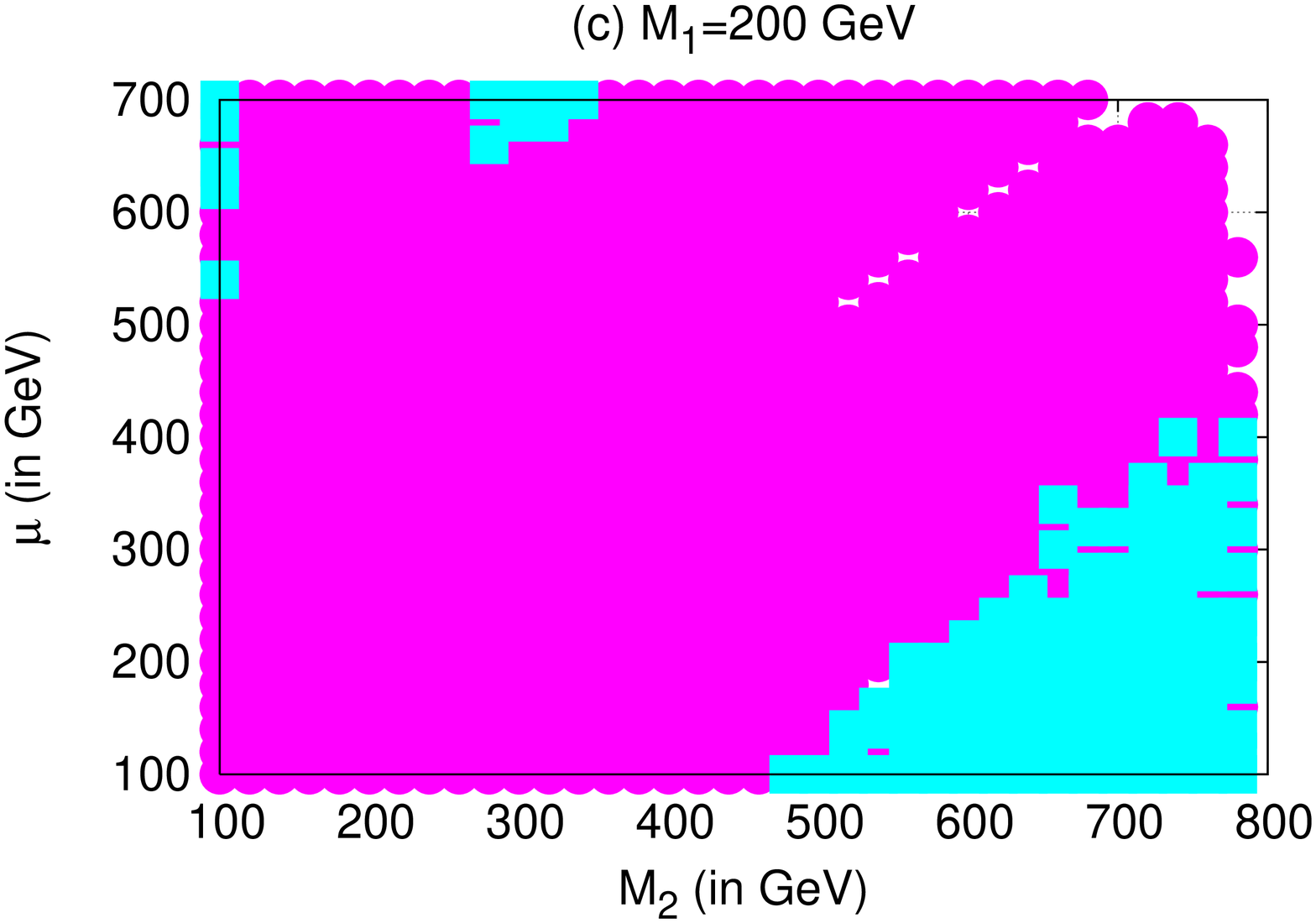,width=5.0 cm,height=5.2cm,angle=-90.0}}	
\caption{The pink (dark) region is the region where the rate for H$^{\pm}$ 
is less than that for $h$ production, for the sky-blue region (grey) 
it is the opposite and white regions indicate zero production for both.
We take $\mone=\mtwo/2$ (a), $\mone=100$ GeV (b) and $\mone=200$ GeV
(c). $m_{H^{\pm}}$ is set to 250 GeV.} 
\end{center}
\label{fig11}
\end{figure}

\section{Conclusions}
We have investigated the rates  for production of different physical
 Higgs states in SUSY cascades at the LHC. The new inputs are 
(a) relaxation of the universality condition relating electroweak
gaugino masses and (b) allowing sleptons to be light enough to be
 produced on-shell in cascades. As we show through the dependence of
 production rates on various parameters, both these inputs can
 significantly affect the
phenomenology at the LHC. The most important observation is that the
presence or absence of gaugino universality is reflected in a rather
interesting complementarity between the relative production rates of
$H^{\pm}$ and $h$. Therefore, one should aim at a careful
identification of charged as well as neutral Higgs signals in
cascades to extract useful information about the underlying SUSY
scenario. A detailed discussion of such signals {\it all} the physical
Higgs states in SUSY and methods of
suppressing their backgrounds will be presented in a separated study. 
\vskip 20pt
{\bf Acknowledgments:}

We acknowledge S. Mrenna for resolving and clarifying some issues in
Pythia and Peter Skands for useful discussions.
Computational work for this study was partially carried out in the cluster
computing facility at Harish-Chandra Research Institute
(http://cluster.mri.ernet.in). This work is partially supported by
Regional Center for Accelerator-based Particle Physics, Harish-Chandra
Research Institute and funded by the Department of Atomic Energy, Government of India under the XIth 5-year Plan.


\end{document}